\documentclass[draftclsnofoot,onecolumn,peerreview]{IEEEtran}
\usepackage{cite}
\usepackage{graphicx}
\usepackage{color}

\begin{document}
\title{Allan variance computed in space domain: Application to InSAR data to characterize noise and geophysical signal}

\author{Olivier Cavali\'{e}
	and~Fran\c{c}ois~Vernotte
\thanks{O. Cavali\'{e} is with Universit\'e de Nice Sophia Antipolis, CNRS, IRD, 
  Observatoire de la C\^ote d'Azur, Géoazur UMR 7329, 250 rue Albert Einstein, Sophia Antipolis 06560 Valbonne, France (Email: \texttt{ocavalie@geoazur.unice.fr}).}
\thanks{F. Vernotte is with UTINAM,
  Observatory THETA of Franche-Comt\'{e},
  University of Franche-Comt\'{e}/CNRS,
41 bis avenue de l'observatoire - B.P. 1615,
25010 Besan\c{c}on Cedex - France (Email: \texttt{francois.vernotte@obs-besancon.fr}).}\\
Version 5 -- \today
}


\maketitle

\begin{abstract}
The Allan variance was introduced fifty years ago for analyzing the stability of frequency standards. Beside its metrological interest, it may be also considered as an estimator of the large trends of the power spectral density (PSD) of frequency deviation. For instance, the Allan variance is able to discriminate different types of noise characterized by different power laws in the PSD. 

The Allan variance was also used in other fields than time and frequency metrology: for more than 20 years, it has been used in accelerometry, geophysics, geodesy, astrophysics and even\ldots finances! However, it seems that up to now, it has been exclusively applied for time series analysis. We propose here to use the Allan variance onto spatial data.

Interferometric synthetic aperture radar (InSAR) is used in geophysics to image ground displacements in space (over the SAR image spatial coverage) and in time thank to the regular SAR image acquisitions by dedicated satellites. The main limitation of the technique is the atmospheric disturbances  that affect the radar signal while traveling from the sensor to the ground and back. In this paper, we propose to use the Allan variance for analyzing spatial data from InSAR measurements. The Allan variance was computed in $XY$ mode as well as in radial mode for detecting different types of behavior for different space-scales, in the same way as the different types of noise versus the integration time in the classical time and frequency application. We found that radial AVAR is the more appropriate way to have an estimator insensitive to the spatial axis and we applied it on SAR data acquired over eastern Turkey for the period 2003-2011. Space AVAR allowed to well characterize noise features, classically  found in InSAR such as phase decorrelation producing white noise or atmospheric delays, behaving like a random walk signal. We finally applied the space AVAR to an InSAR time series to detect when the geophysical signal, %
here the ground motion,  emerges from the noise. 
\end{abstract}

\newpage

\tableofcontents

\newpage

\section{Introduction}
The Allan variance was designed by David Allan 50 years ago for characterizing the time stability of atomic clocks \cite{allan1966}. This statistical tool quickly became a standard for researchers, engineers and manufacturers in the time and frequency field, to the extent that the Allan variance is not only considered as a stability estimator but as the stability itself! The use of this tool then spread to related fields such as inertial sensors \cite{elsheimy2008}, temperature \cite{weng2014} and voltage \cite{witt2005}. Since 2000's, the Allan variance has been used in many other domain including geodesy, to measure absolute displacement of geodetic stations \cite{lebail2004}, astrophysics, for selecting stable extragalactic compact radio sources from the VLBI program \cite{feissel2003}, geophysics, by monitoring the elevation of a volcano station to detect a change in the Allan deviation signature as a symptom of an eruption threat \cite{roberts2002}. The Allan variance is even used in finance for optimizing the financial returns \cite{hernandez2012}! However, all these applications of the Allan variance deal with time series analysis. To our knowledge, this statistical estimator was never used over two-dimensional spatial data. This is the aim of this paper: modifying the 1-D Allan variance to a 2-D space Allan variance for fulfilling the needs of geophysics in terms of Earth surface deformation monitoring.  

Earth surface deformation is one of the most important observation to infer earth interior properties  \cite{kaufmann00a,piersanti99,pollitz00}, subduction zone activity \cite{cavalie13} or natural hazard, including seismic hazard,  landslide \cite{colesanti06} or land subsidence \cite{schmidt03}. While looking at large wavelength phenomena, measuring accurately how Earth deforms is a challenging task. A very effective way has been developed  with geodetic techniques such as Global Navigation Satellite System (GNSS) or radar interferometry (InSAR). In both cases, %
measurements use the travel time of electromagnetic waves between satellites and Earth's surface to derive the ground displacements. As Earth's atmosphere is not neutral, taking into account the wave travel in this media is a key parameter %
to get reliable measurement of surface deformation. 

Here, we investigate the noise signature in InSAR and in particular the noise induced by the wave propagation into the atmosphere. InSAR has proven to be a very powerful tool to measure Earth surface displacements \cite{simons07}. %
Despite it measures the deformation in 1 dimension, this is the only technique offering  large spatial coverage (hundreds of kilometers) and excellent spatial resolution (tens of meters). The main limitation is the atmospheric delays during the wave propagation in the atmosphere. Indeed, spatiotemporal variations of the air refractivity induce additional signal on the interferograms, that we called atmospheric phase screen (APS). Following atmospheric conditions during the SAR images acquisitions, APS on interferograms can reach up to 10 cm, or more in case of extreme meteorological events. It is thus potentially much larger than the geophysical signal. A method to model the structure and, eventually, correct the APS is then an important goal to fully exploit InSAR as an accurate method for displacement measurement. 
The extension of AVAR to 2-dimensional images could be a powerful tool for achieving this purpose.

\section{Choice of the variance}
	\subsection{Expected properties\label{sec:exp_prop}}
The Allan variance is designed for assessing the large trends of the spectrum, not for detecting fine spectral lines. On the other hand, it can easily distinguish different power-laws in the spectrum of a time signal. This is exactly what we need in the analysis of InSAR images. 

\begin{figure}
\includegraphics[width=8.5cm]{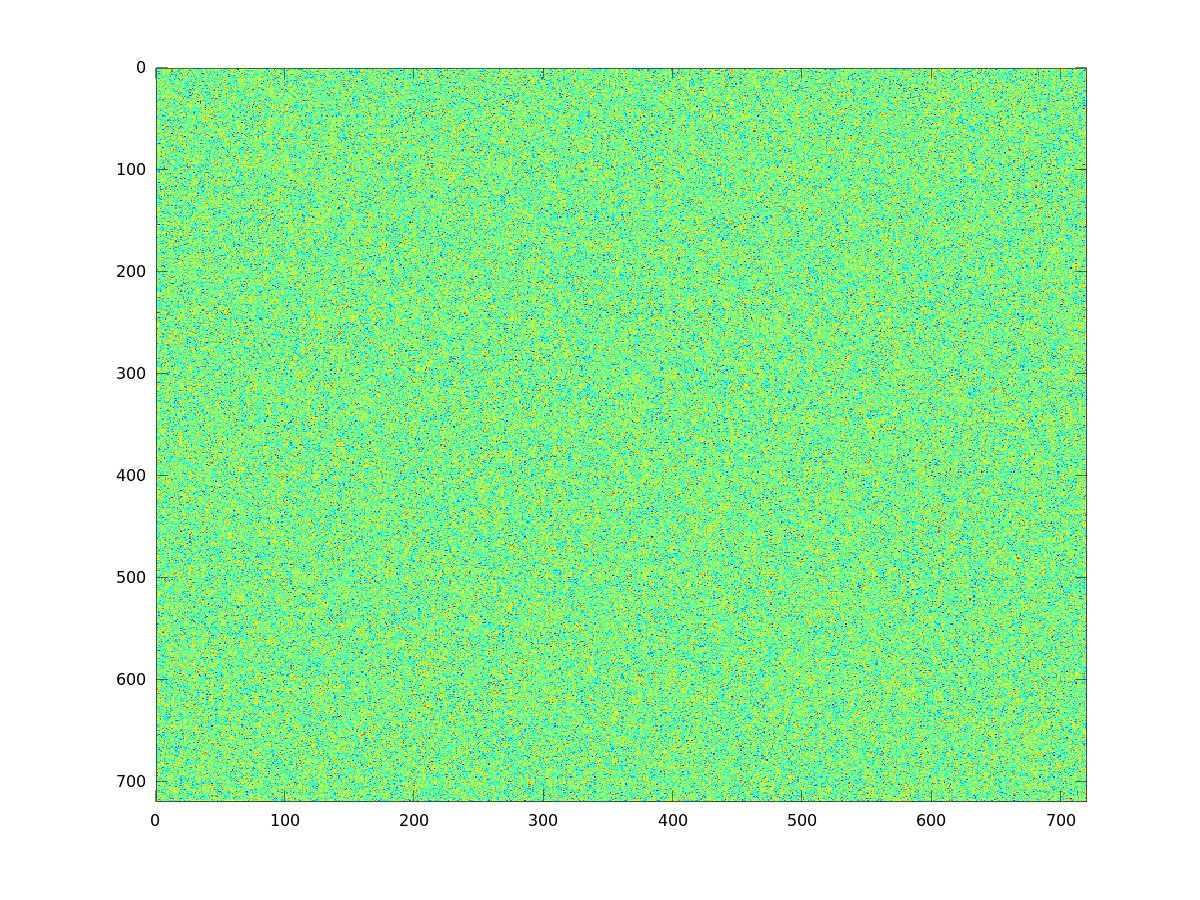} \hfill \includegraphics[width=8.5cm]{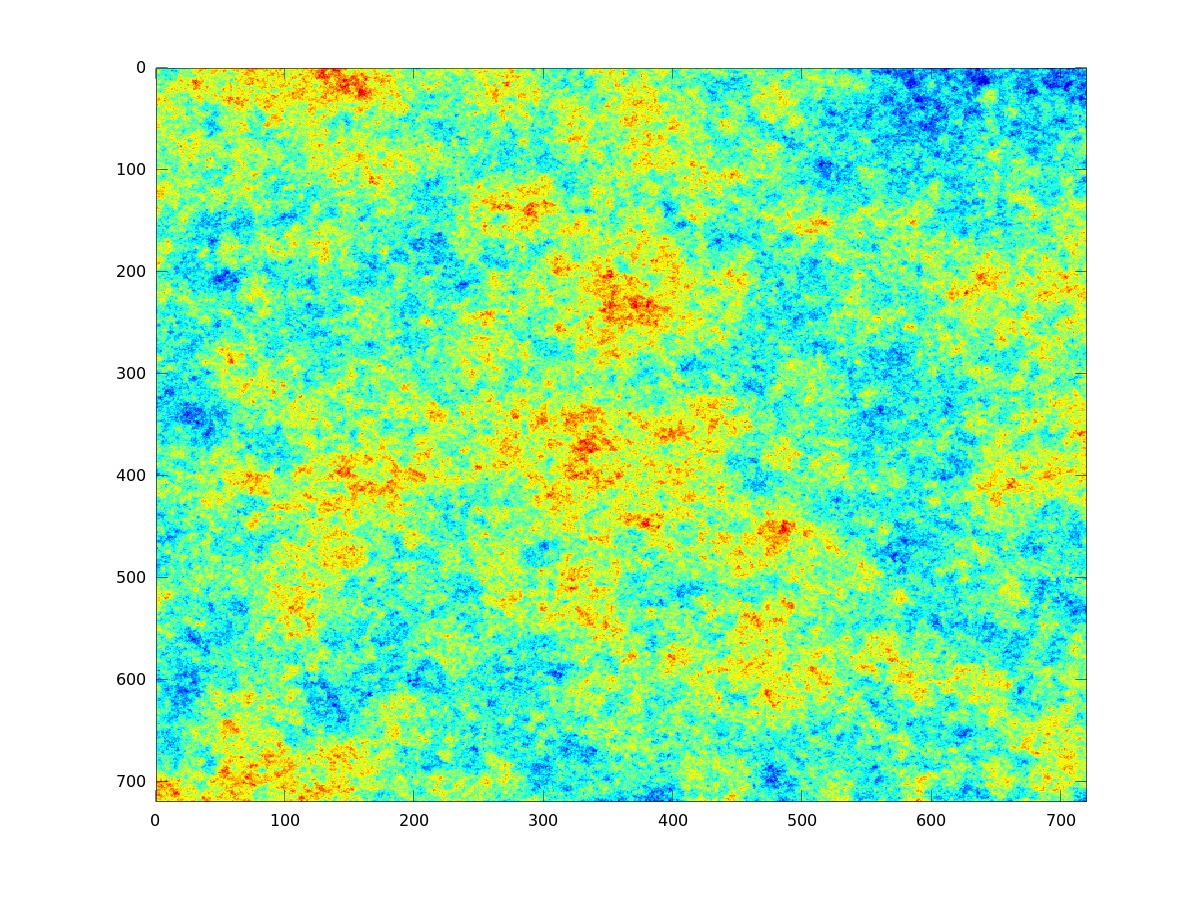}
\caption{Example of white noise (left) and random walk (right).\label{fig:exnoise}}
\end{figure}

We expect that the noise may be modeled as a power law of the wave number $k$, where $k$ is the dual variable of the radius in polar coordinate, exactly as the frequency $f$ is the dual variable of the time $t$ in time series. Like in time and frequency, we could encounter a Gaussian white noise, of electronic origin or due to phase decorrelation, 
and an atmospheric noise exhibiting large scale correlation, likely with a power spectrum decreasing as $k^{-2}$ (random walk). Figure \ref{fig:exnoise} displays these two types of noise. Ideally, the response of Allan variance should be a plane with different slopes, depending on the type of noise: horizontal for the atmospheric noise and decreasing with the scale factor for white noise. 

\begin{figure}
\includegraphics[width=8.5cm]{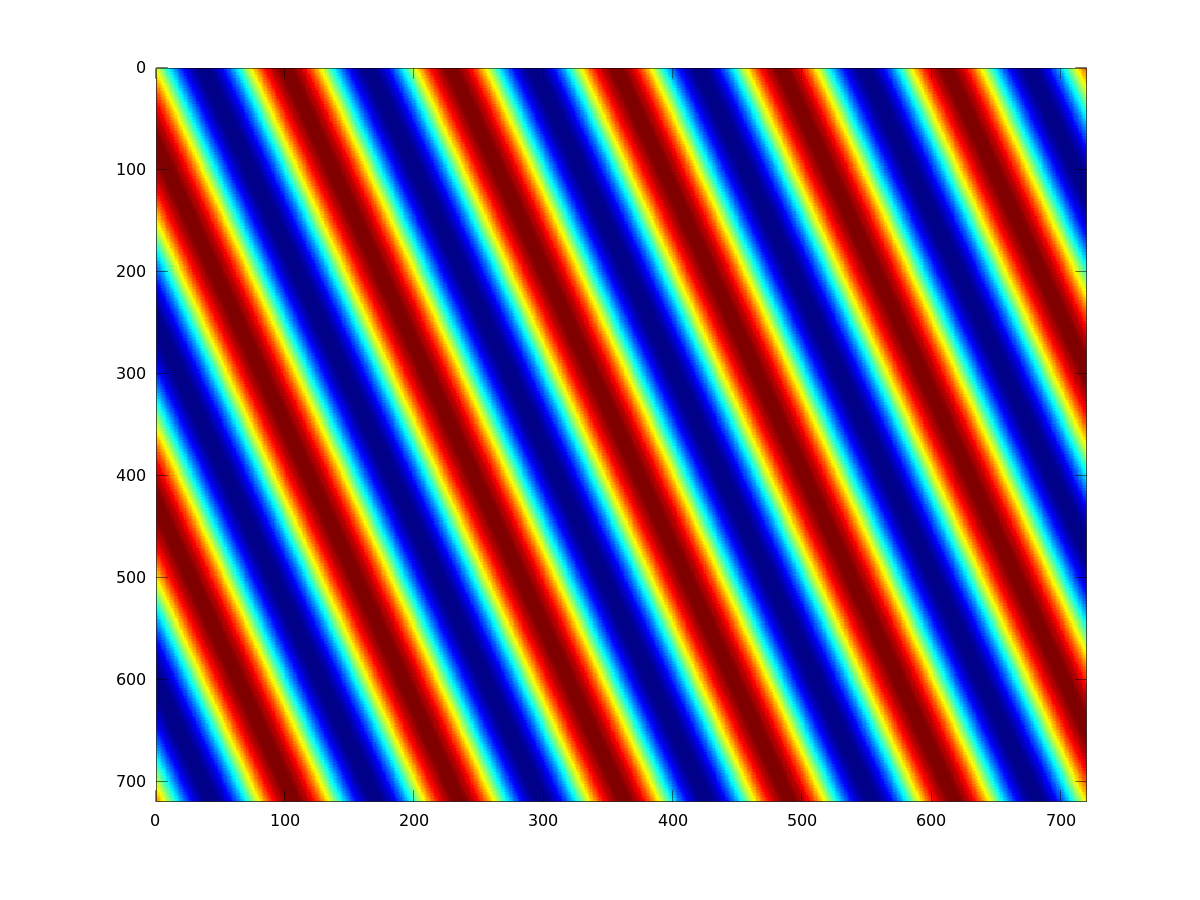} \hfill \includegraphics[width=8.5cm]{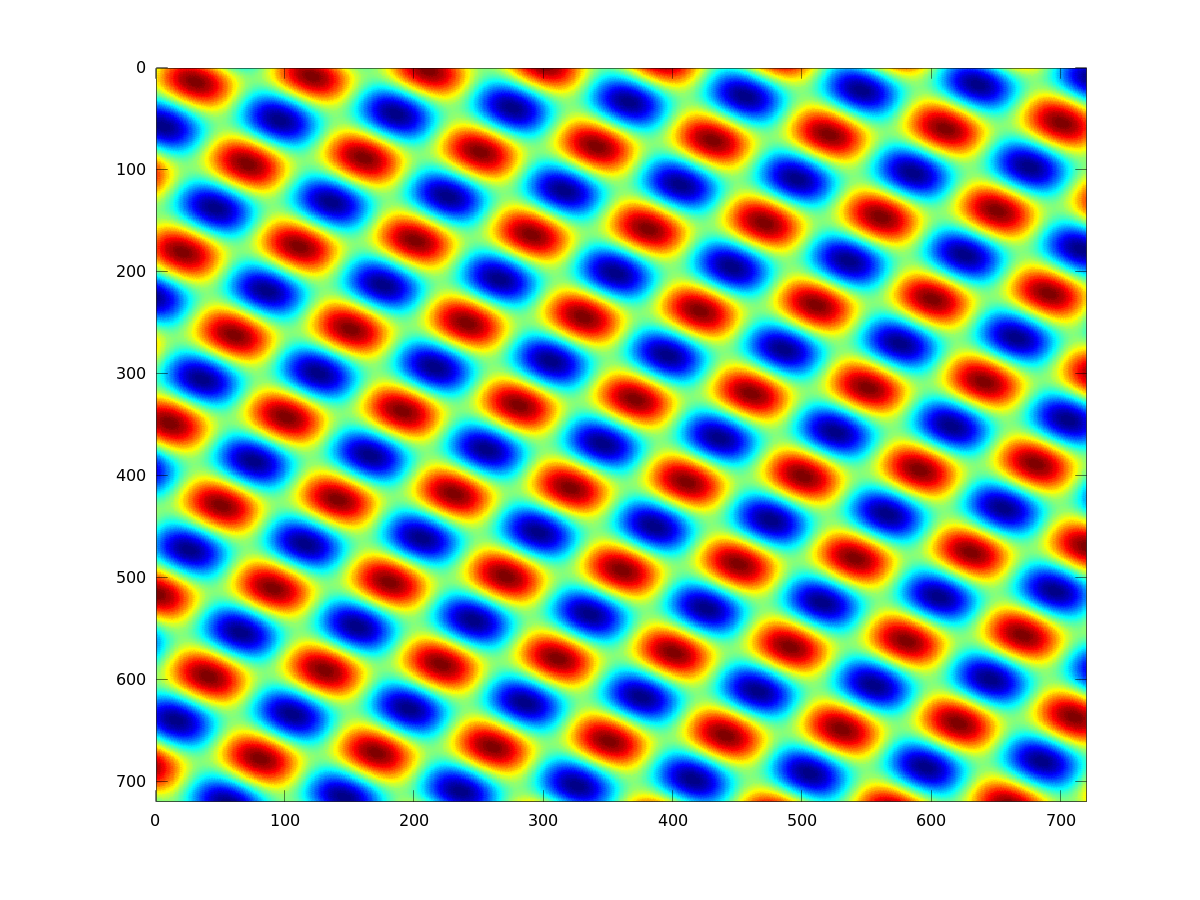}
\caption{Example of an 1-dimensional sine image (left) and of 2-dimensional sine image (right).\label{fig:sinpic}}
\end{figure}

This behavior should permit to separate the noise signature from the signal signature which depends on the topography of the images. The Allan variance must detect the large structure of the image signal. We ought then to identify the response of space AVAR\footnote{Since the classical Allan variance is often denoted AVAR, we will refer to the 2-dimensional space Allan variance as \textit{space AVAR}.} for various features such as linear slope, a 1-dimensional sine, a 2-dimensional sine, with different orientations. Such patterns may be seen on Figure \ref{fig:sinpic}.

Finally, on a time series of images covering the same area %
site over several years, we should be able to see the signal emerging from the noise as the date of the image increases. 

	\subsection{$XY$-AVAR}
\begin{figure}
\begin{minipage}[c]{5cm}
\includegraphics[width=5cm]{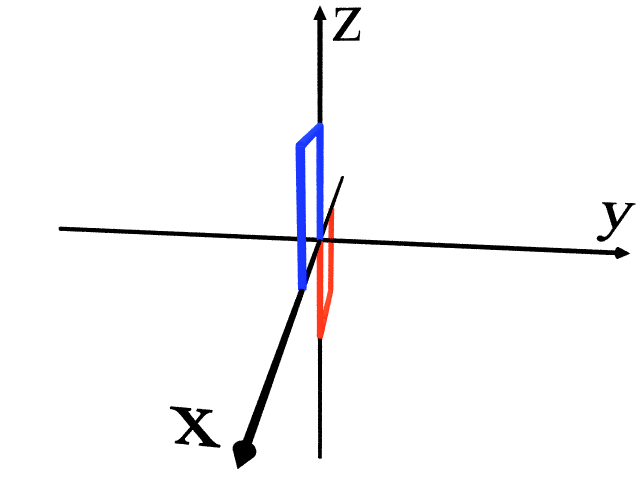}
\end{minipage}
$\times$
\begin{minipage}[c]{5cm}
\includegraphics[width=5cm]{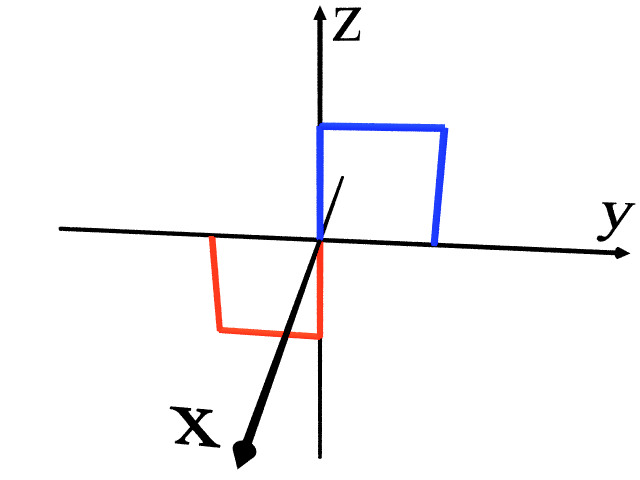}
\end{minipage}
$=$
\begin{minipage}[c]{5cm}
\includegraphics[width=5cm]{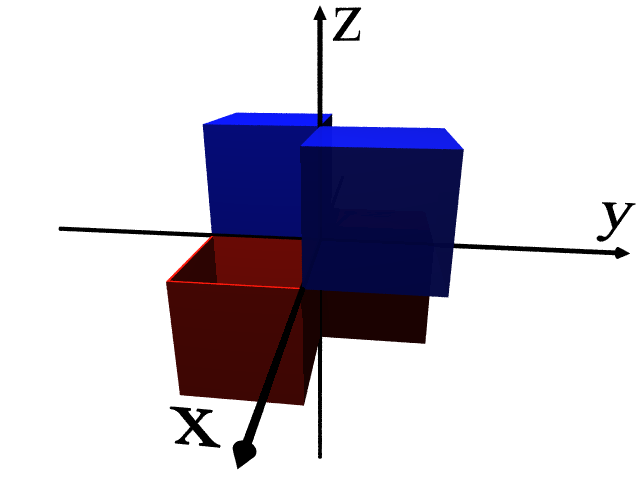}
\end{minipage}
\caption{From classical 1D Allan variance to $XY$-AVAR.\label{fig:AVARX2XY}}
\end{figure}

The first attempt for extending the classical Allan variance from 1D to 2D was the $XY$-AVAR. This variance relies upon a duplication of the double-stepped shape of AVAR along the first dimension, i.e. $X$-axis, in the second dimension, i.e. $Y$-axis (see Figure \ref{fig:AVARX2XY}).

		\subsubsection{Definition}
	\paragraph{Kernel} If $XY$-AVAR is considered as a 2D-filter, the kernel is its impulse response $h_{XY}(x,y)$ which is the product of the kernel of classical AVAR along $X$ by the kernel of classical AVAR along $Y$. Let us denote $h_X(x)$ the kernel of AVAR along $X$: 
\begin{equation}
\left\{\begin{array}{lcl}
\displaystyle h_X(x)=-\frac{1}{\sqrt{2} \lambda_X} & \textrm{if} & -\lambda_X \leq x < 0\\
\displaystyle h_X(x)=+\frac{1}{\sqrt{2} \lambda_X} & \textrm{if} & 0 \leq x < +\lambda_X\\
h_X(x)=0 & \textrm{if} & x < -\lambda_X \quad \textrm{or} \quad x \geq +\lambda_X
\end{array}\right.
\end{equation}
where $\lambda_X$ is the scale factor along $X$.

Similarly, let us denote $h_Y(y)$ the kernel of AVAR along $Y$: 
\begin{equation}
\left\{\begin{array}{lcl}
\displaystyle h_Y(y)=-\frac{1}{\sqrt{2} \lambda_Y} & \textrm{if} & -\lambda_Y \leq y < 0\\
\displaystyle h_Y(y)=+\frac{1}{\sqrt{2} \lambda_Y} & \textrm{if} & 0 \leq y < +\lambda_Y\\
h_Y(y)=0 & \textrm{if} & x < -\lambda_Y \quad \textrm{or} \quad y \geq +\lambda_Y
\end{array}\right.
\end{equation}
where $\lambda_Y$ is the scale factor along $Y$.

The kernel of $XY$-AVAR is then:
\begin{equation}
h_{XY}(x,y)=h_X(x)\cdot h_Y(y).
\end{equation}

	\paragraph{Transfer function} The transfer function is the Fourier transform of the kernel (i.e. the impulse response). Moreover, since the kernel is the product of one function depending only on $x$ by a function depending only on $y$, its Fourier transform is the product of the Fourier transform of these functions (see \cite{roddier1978} p. 139):
\begin{equation}
H_{XY}(u,v)=H_X(u) \cdot H_Y(v)
\end{equation}
where $u$ and $v$ are spatial frequencies in $X$ and $Y$ directions.

Since the squared modulus of the transfer function of the classical Allan variance is given by \cite{rutman1978}:
\begin{equation}
|H_{AVAR}(f)|^2=2\frac{\sin^4(\pi \tau f)}{(\pi \tau f)^2}
\end{equation}
the squared modulus of the transfer function of $XY$-AVAR is then:
\begin{equation}
|H_{XY}(u,v)|^2=4\frac{\sin^4(\pi \lambda_X u)}{(\pi \lambda_X u)^2}\cdot \frac{\sin^4(\pi \lambda_Y v)}{(\pi \lambda_Y v)^2}.
\end{equation}

\begin{figure}
\centering \includegraphics[width=10cm]{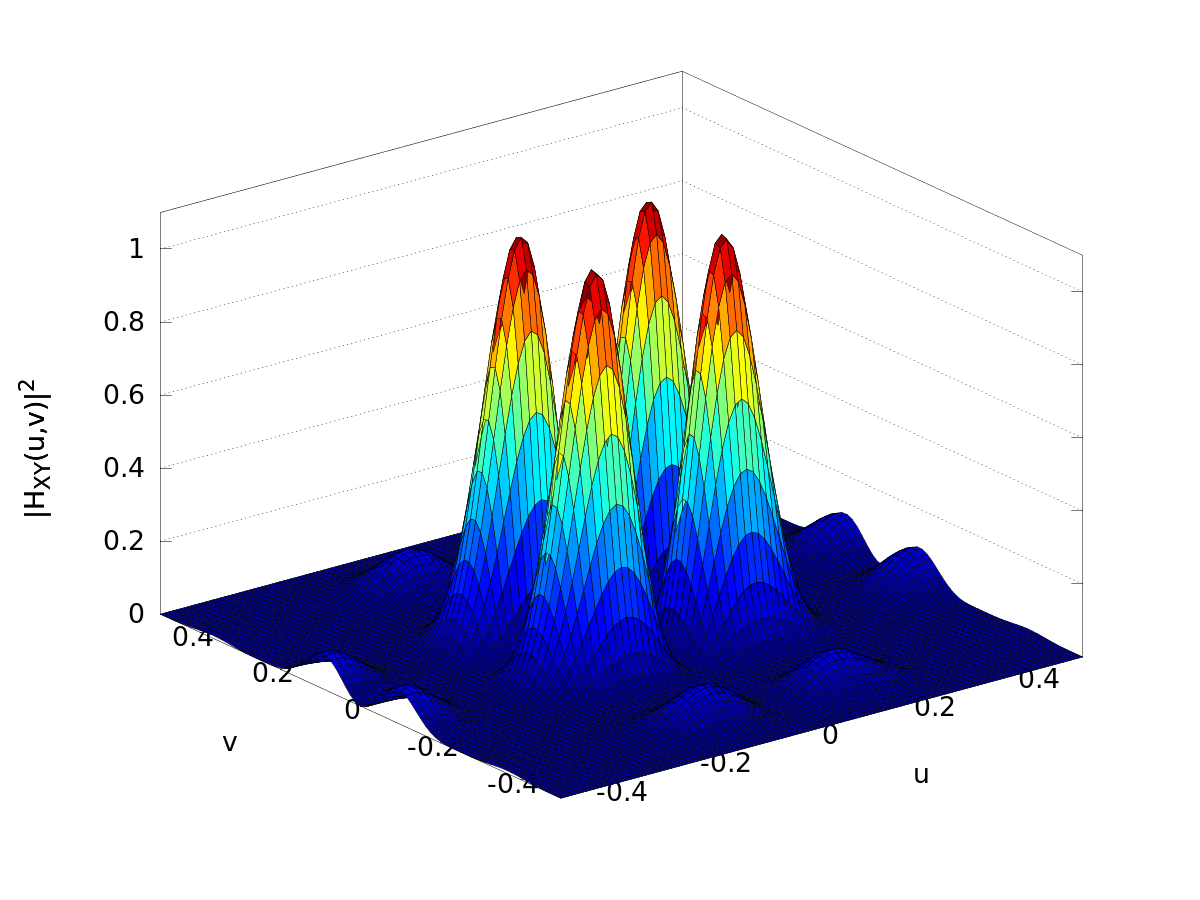}
\caption{Squared modulus of the transfer function of $XY$-AVAR for $\lambda_X=4$ pixels and $\lambda_Y=3$ pixels.\label{fig:trfuXY}}
\end{figure}

	\paragraph{First feeling\label{sec:ffXY}} Figure \ref{fig:trfuXY} shows the shape of this transfer function. This graph exhibits a grid of zeros since each time one of the functions $H_X(u)$ or $H_Y(v)$ is equal to 0, it yields a line of 0 along the perpendicular axis. Therefore, $XY$-AVAR is strongly dependent on the axis orientation. This is clearly a disadvantage because the response of $XY$-AVAR may vary significantly versus the rotation of the image, although the image content %
 is obviously the same.
		\subsubsection{Responses of $XY$-AVAR}
\begin{figure}
\includegraphics[width=8.5cm]{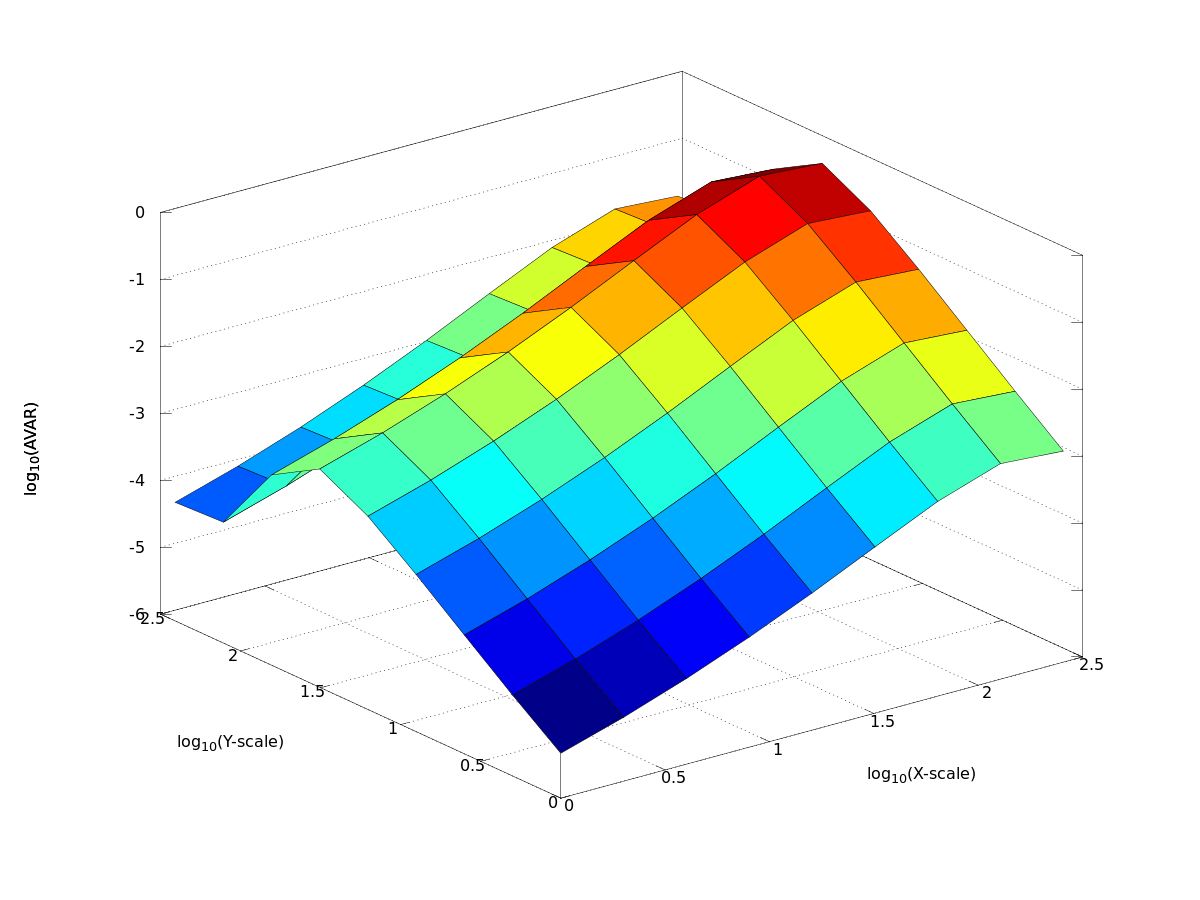} \hfill \includegraphics[width=8.5cm]{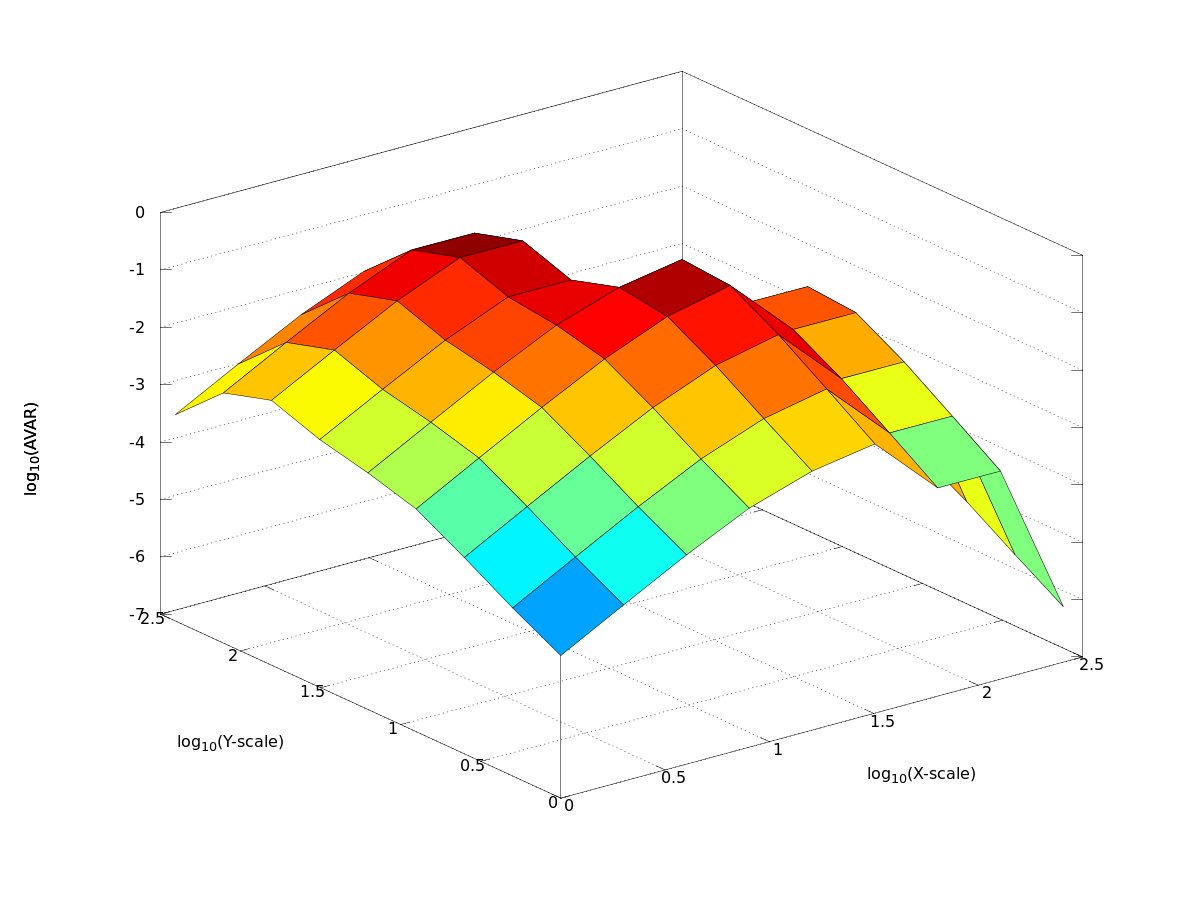}
\caption{$XY$-AVAR of a 1-D sine (left) and of a 2-D sine (right).\label{fig:XYsine}}
\end{figure}

In order to test $XY$-AVAR, we applied it to the various noises and geometrical patterns defined in \S \ref{sec:exp_prop}.
			\paragraph{Response to a 1-D sine}
The response of $XY$-AVAR to 1-D sine depends drastically on its orientation: if the sine is exactly in the direction of one axis, the $XY$-AVAR response is null; but this is no longer the case when the 1-D sine orientation is arbitrary as shown on Figure \ref{fig:XYsine} (left).%

			\paragraph{Response to a 2-D sine}
Similarly, if the 2-D sine are in the direction of the axis, $XY$-AVAR yields a null response. Otherwise, we can see an example of  $XY$-AVAR response %
on Figure \ref{fig:XYsine} (right).

			\paragraph{Response to a drift} In this case also, the response may be null if the drift is oriented in the direction of one of the axis.

\begin{figure}
\includegraphics[width=8.5cm]{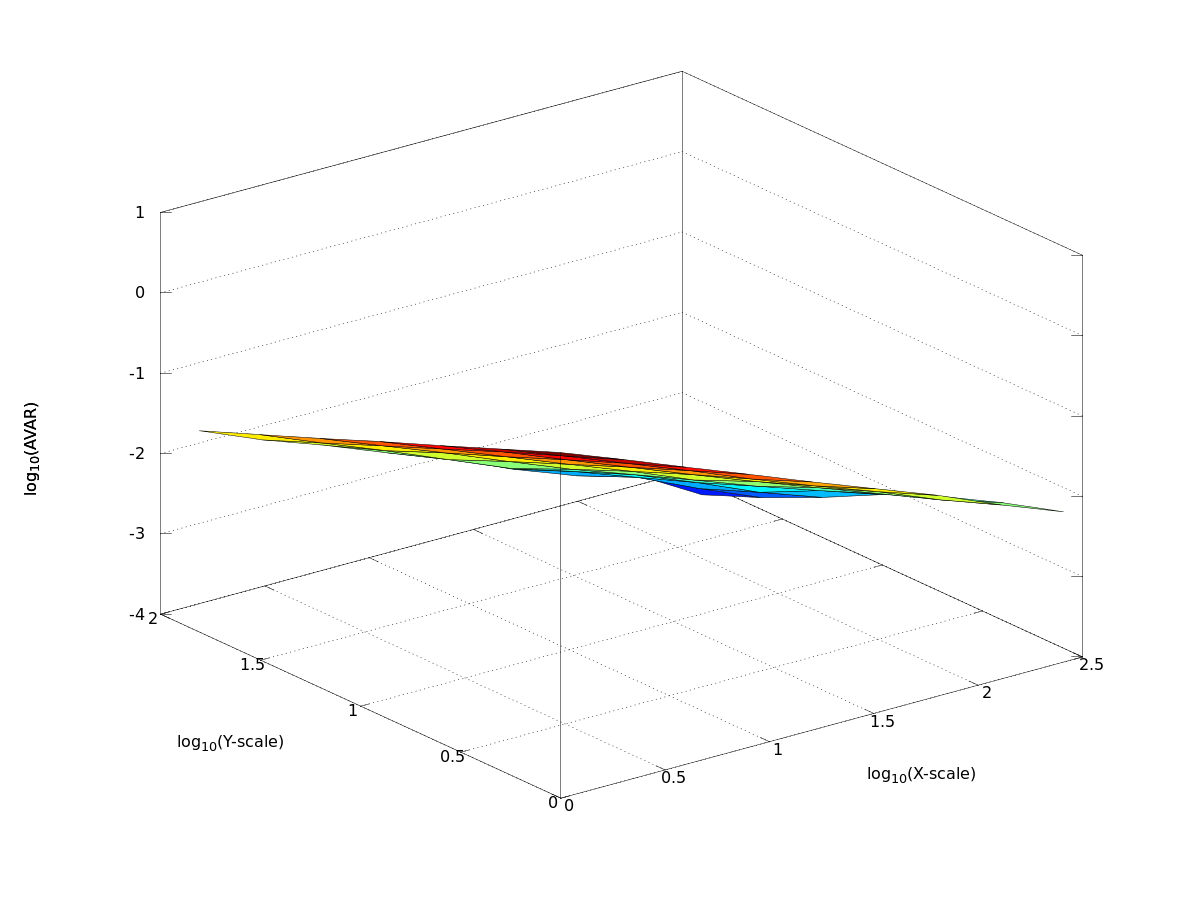} \hfill \includegraphics[width=8.5cm]{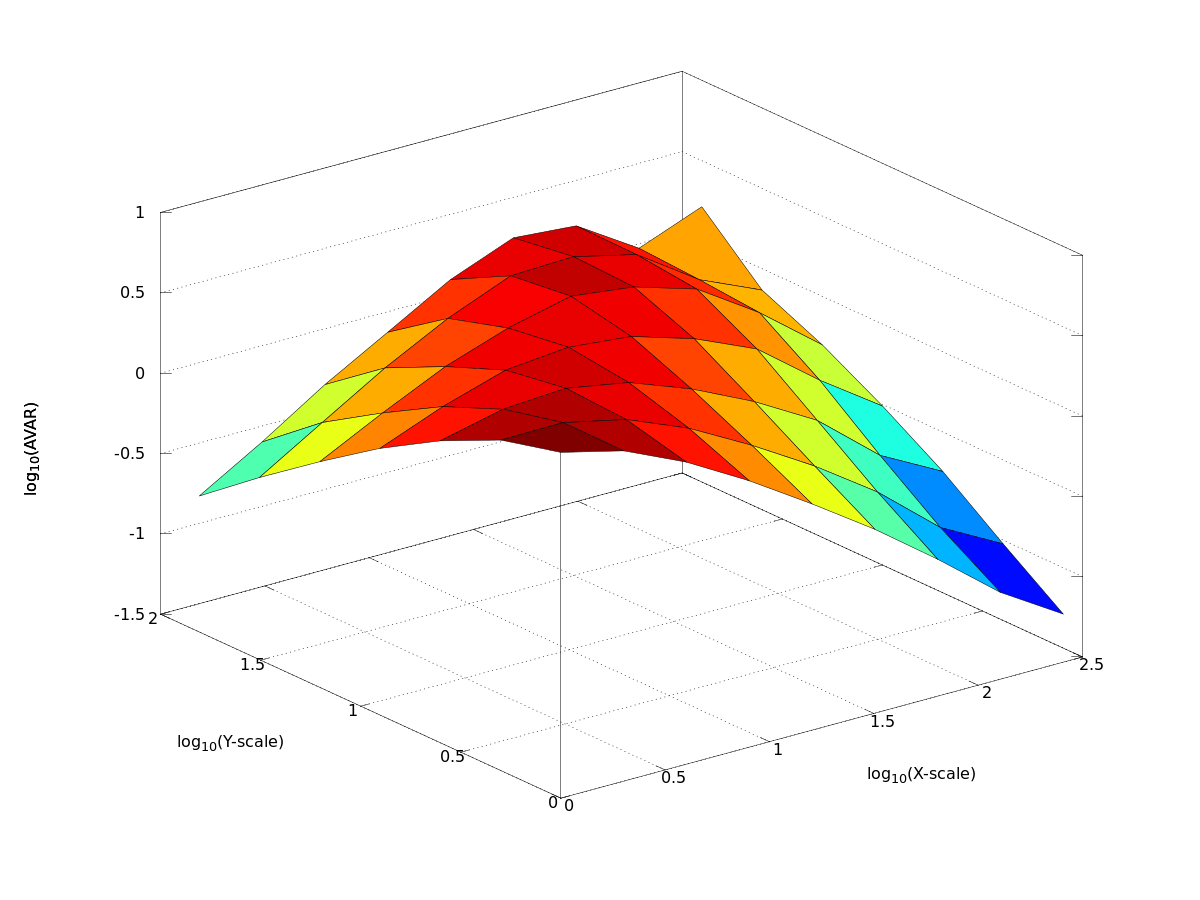}
\caption{$XY$-AVAR of a white noise (left) and of a random walk (right).\label{fig:XYnoises}}
\end{figure}

			\paragraph{Response to a Gaussian white noise}
As expected, the response of $XY$-AVAR for a white noise is a plan decreasing proportionally to the scale factor. Figure \ref{fig:XYnoises} (left) shows this plane seen almost edge-on.

			\paragraph{Response to a random walk}
On the contrary, the response of $XY$-AVAR for a random walk is not at all flat (see Figure \ref{fig:XYnoises}, right): its curvature spans over 2 decades. Moreover, %
the surface exhibits a protruding central diagonal with flanged edges. This feature is not suitable because it could interfere with the signal signature.
 
\begin{figure}
\includegraphics[width=10cm]{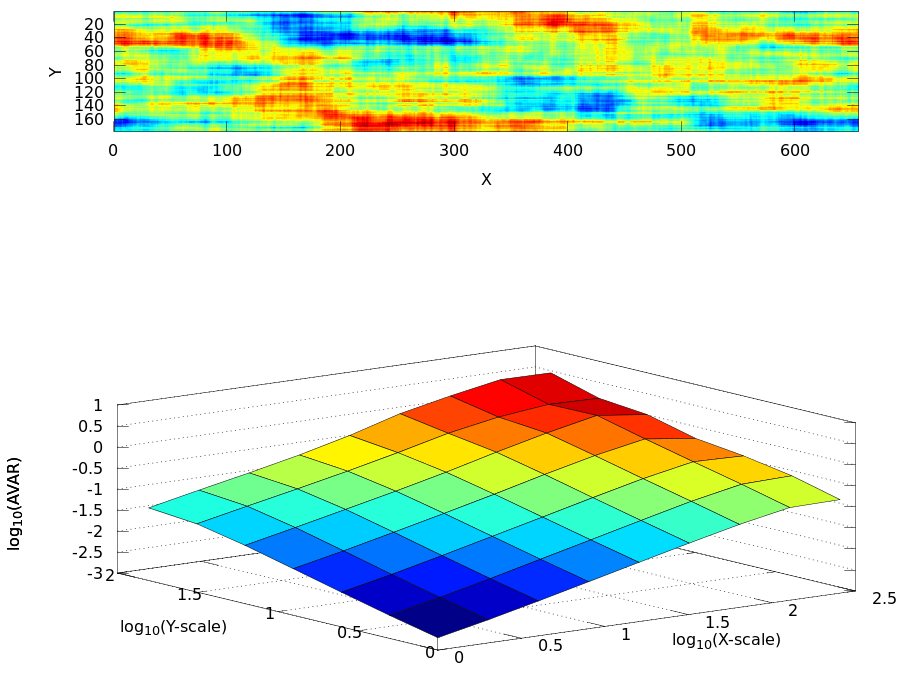}
\caption{Example of $u^{-2}v^{-2}$ noise (above) and the response of $XY$-AVAR for this noise (below).\label{fig:XYRW}}
\end{figure}

			\paragraph{Searching for a noise giving a flat horizontal response}
The only noise which could give a flat horizontal response by $XY$-AVAR must have a power spectrum which decreases as $u^{-2}v^{-2}$ rather than $k^{-2}$ as a classical random walk. Figure \ref{fig:XYRW} shows such a noise and the response of $XY$-AVAR for this noise. Firstly, we can see that this response is flat but not horizontal: it increases as the scale factor. Secondly, the response is only flat if the axis of the image are aligned to the axis of $XY$-AVAR. Moreover, the image of this noise is not at all natural: the $X$ and $Y$ axis are very marked as a raster image.

		\subsubsection{Conclusion about $XY$-AVAR}
As expected in section \S \ref{sec:ffXY}, $XY$-AVAR is not suitable because it is too much dependent on its axis orientation. This has two consequences:
\begin{itemize}
	\item some responses (for a drift, for a 1-D or 2-D sine) may completely vanish depending on the axis orientation
	\item a random walk does not yield a flat and horizontal response but a complex feature which could be confused with the response for the signal.
\end{itemize}
For these reasons, another space Allan variance algorithm must be sought.

	\subsection{Radial AVAR: Allan's hat}
\begin{figure}
\centering
\begin{minipage}[c]{5cm}
\includegraphics[width=5cm]{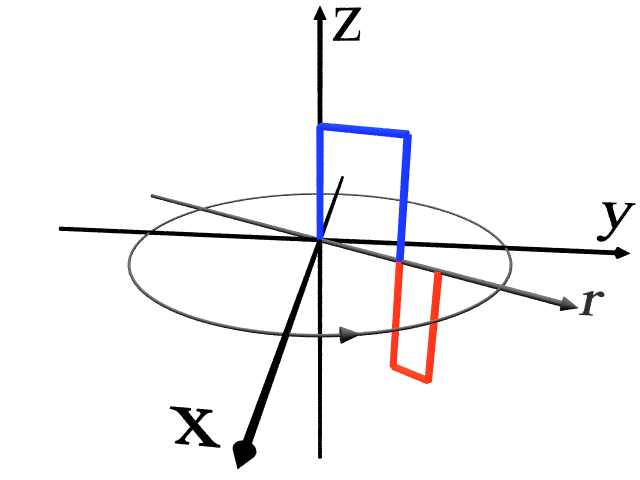}
\end{minipage}
 $\longrightarrow$ 
 \begin{minipage}[c]{5cm}
\includegraphics[width=5cm]{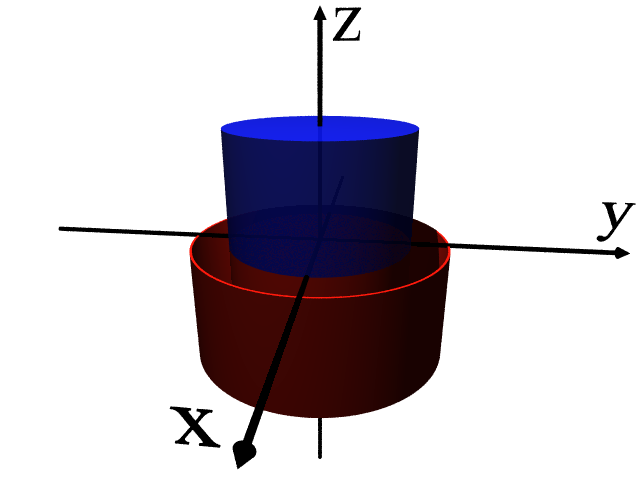}
\end{minipage}
\caption{From classical 1D Allan variance to Allan's hat.\label{fig:allan-s_hat}}
\end{figure}

We ought then to find a variance which does not depend too much on the orientation of the axis. Thus, we chose to define a kernel with a radial symmetry. The principle is to rotate the double-stepped shape of AVAR around the $Z$-axis (see Figure \ref{fig:allan-s_hat}). In order to preserve the same surface for the central positive plateau and the peripheral negative one, it is necessary to respect a condition on the radii. %
Let us denote $r_1$, the radius of the central positive step, $r_2$, the radius of the peripheral negative step, $S_1$ the surface of the central positive plateau, and $S_2$, the surface of the peripheral negative plateau:
\begin{equation}
S_1=\pi r_1^2=S_2=\pi \left(r_2^2-r_1^2\right) \quad \Rightarrow \quad r_2=\sqrt{2} r_1.
\end{equation}
This leads to a 2-D shape %
that resembles a hat which we named the ``\textit{Allan's hat}''.
		\subsubsection{Definition}
	\paragraph{Kernel} Considering the full radial symmetry of Allan's hat, the kernel may be written in polar coordinates as:
\begin{equation}
\left\{\begin{array}{lcl}
\displaystyle h_\mathrm{rad}(r,\theta)=+\frac{1}{\sqrt{2}\pi r_1^2} &\textrm{if}& 0\leq r<r_1\\
\displaystyle h_\mathrm{rad}(r,\theta)=-\frac{1}{\sqrt{2}\pi r_1^2} &\textrm{if}& r_1\leq r<\sqrt{2}r_1\\
h_\mathrm{rad}(r,\theta)=0 &\textrm{if}& r\geq \sqrt{2}r_1
\end{array}\right.\label{eq:radavarsym}
\end{equation}
where $r=\sqrt{x^2+y^2}$ and $\theta=\arctan{y/x}$. Note that the factor $1/\sqrt{2}$ has been kept for the sake of homogeneity with the definition of classical AVAR. As defined by (\ref{eq:radavarsym}), radial AVAR is fully insensitive to $\theta$. However, we must be able to detect %
the variations versus the scale factor $\lambda_X$ along $X$-axis as well as the variations versus the scale factor $\lambda_Y$ along $Y$-axis. For this purpose, instead of $r$, we define $\rho$ as:
\begin{equation}
\rho=\sqrt{\left(\frac{x}{\lambda_X}\right)^2+\left(\frac{y}{\lambda_Y}\right)^2}.
\end{equation}
The Allan'hat is now elliptical: the semi-major axis of the central plateau is $\max\left\{\lambda_X,\lambda_Y\right\}$, its semi-minor axis is  $\min\left\{\lambda_X,\lambda_Y\right\}$, its surface is $S_1=\pi\lambda_X\lambda_Y$, and its bounds are given by $\rho<1$. We can now rewrite the kernel in Cartesian coordinates:
\begin{equation}
\left\{\begin{array}{lcl}
\displaystyle h_\mathrm{rad}(x,y)=+\frac{1}{\sqrt{2}\pi \lambda_X \lambda_Y} &\textrm{if}& \displaystyle 0\leq \left(\frac{x}{\lambda_X}\right)^2+\left(\frac{y}{\lambda_Y}\right)^2<1\\
\displaystyle h_\mathrm{rad}(x,y)=-\frac{1}{\sqrt{2}\pi \lambda_X \lambda_Y} &\textrm{if}& \displaystyle 1\leq \left(\frac{x}{\lambda_X}\right)^2+\left(\frac{y}{\lambda_Y}\right)^2<2\\
h_\mathrm{rad}(x,y)=0 &\textrm{if}& \displaystyle \left(\frac{x}{\lambda_X}\right)^2+\left(\frac{y}{\lambda_Y}\right)^2\geq 2.
\end{array}\right.
\end{equation}
	\paragraph{Transfer function} In optics, it is well known that the Fourier transform of a circular aperture of transparency 1 and of radius 1 is $2 J_1(k)/k$ (the Airy pattern) where $J_1(k)$ is the Bessel function of first kind and order 1 and $k$ the wave number (see for example \cite{born1970}, pp. 395--396). Therefore, the squared modulus of the transfer function of radial AVAR is:
\begin{equation}
|H_\mathrm{rad}(k)|^2=\frac{2}{(\pi\lambda_X\lambda_Y)^2}\left[\frac{J_1(k)}{k}-\frac{J_1(\sqrt{2}k)}{\sqrt{2}k}\right]^2.\label{eq:trfuradz}
\end{equation}
Let us define the frequency variables $u$ and $v$ in such a way that $k=\sqrt{\lambda_X^2 u^2 + \lambda_y^2 v^2}$. We can rewrite (\ref{eq:trfuradz}) as:
\begin{equation}
|H_\mathrm{rad}(u,v)|^2=\frac{2}{(\pi\lambda_X\lambda_Y)^2}\left[\frac{J_1\left(\sqrt{\lambda_X^2 u^2 + \lambda_y^2 v^2}\right)}{\sqrt{\lambda_X^2 u^2 + \lambda_y^2 v^2}}-\frac{J_1\left(\sqrt{2(\lambda_X^2 u^2 + \lambda_y^2 v^2)}\right)}{\sqrt{2(\lambda_X^2 u^2 + \lambda_y^2 v^2)}}\right]^2.
\end{equation}

\begin{figure}
\centering \includegraphics[width=10cm]{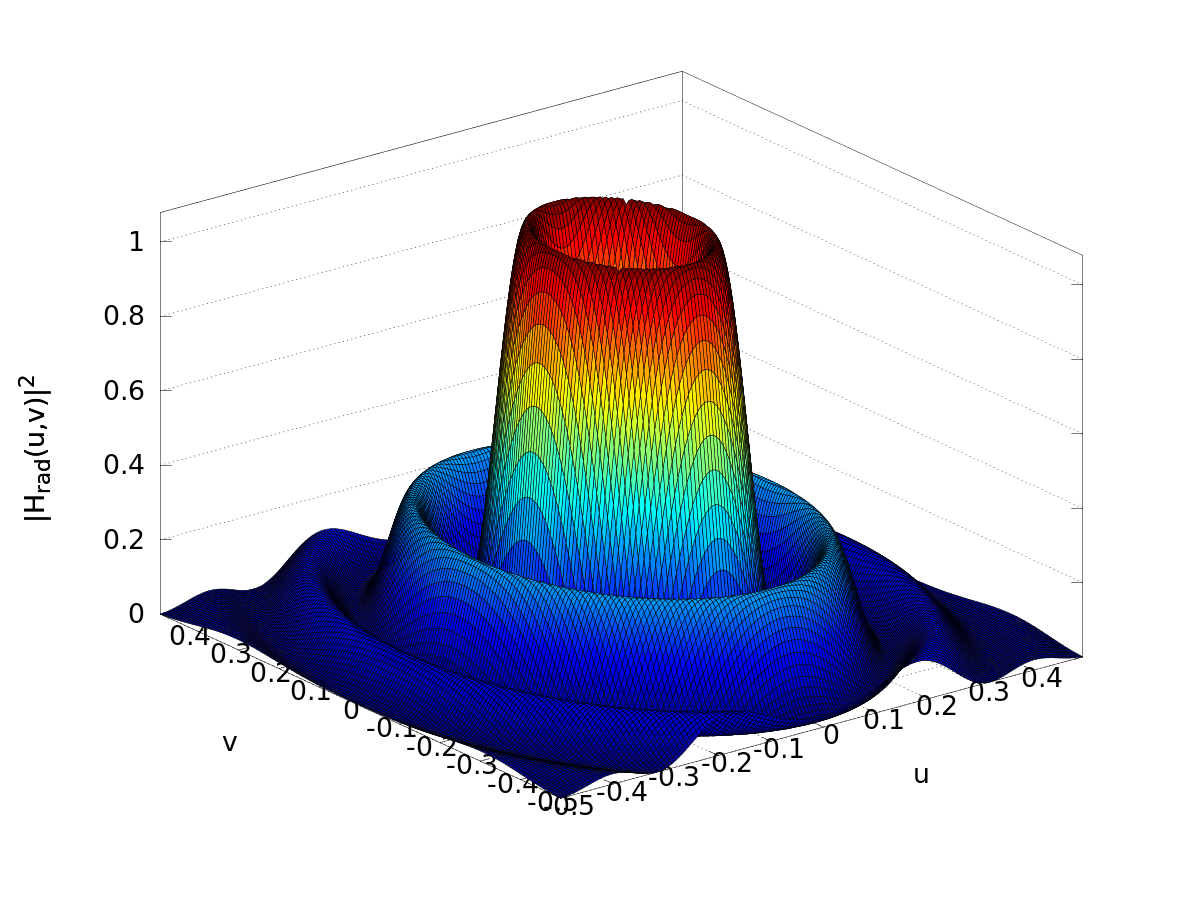}
\caption{Squared modulus of the transfer function of rad-AVAR for $\lambda_X=8$ pixels and $\lambda_Y=6$ pixels.\label{fig:trfurad}}
\end{figure}

The shape of this transfer function may be seen on Figure \ref{fig:trfurad}.
	\paragraph{First feeling} Rad-AVAR seems to be a promising algorithm. Let us study its responses for the various noises and geometrical patterns defined in \S \ref{sec:exp_prop}.
		\subsubsection{Responses}
\begin{figure}
\includegraphics[width=8.5cm]{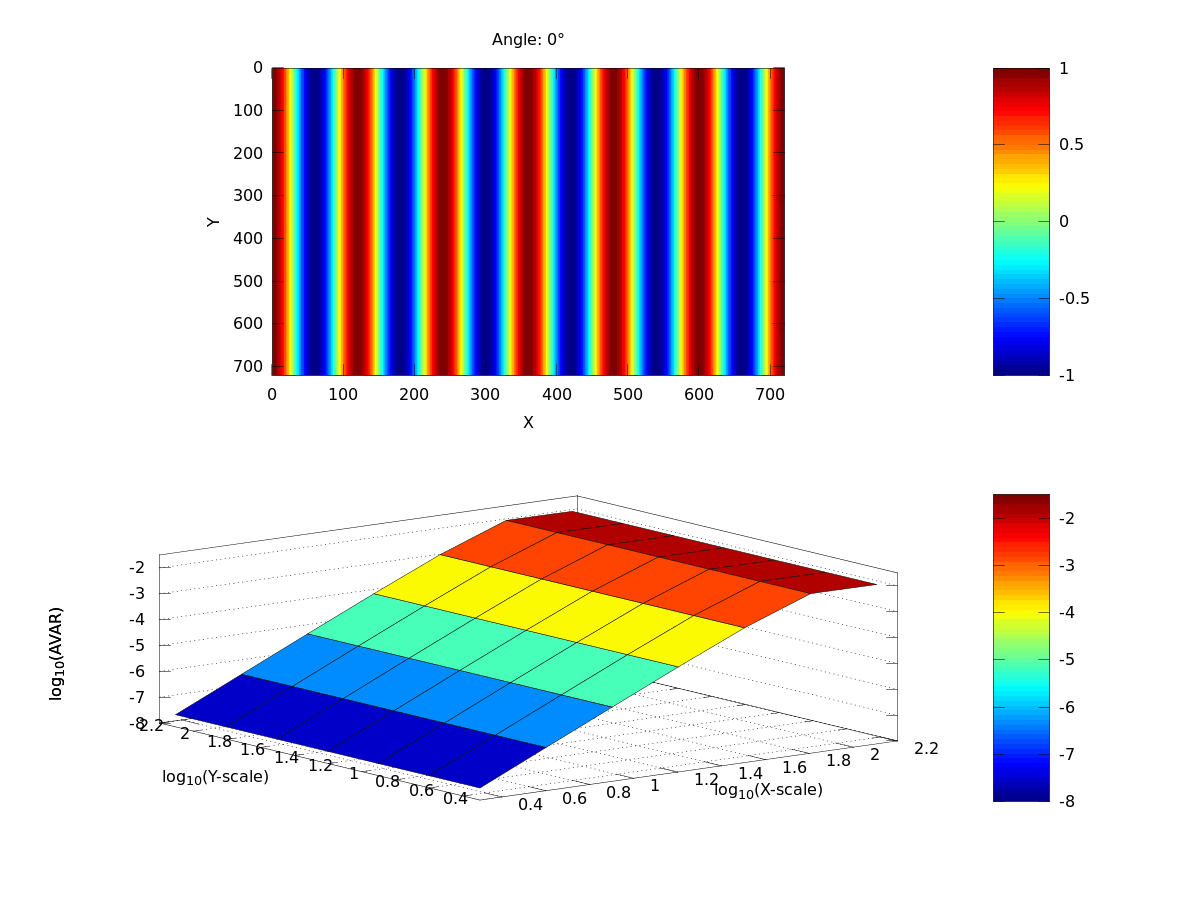} \hfill \includegraphics[width=8.5cm]{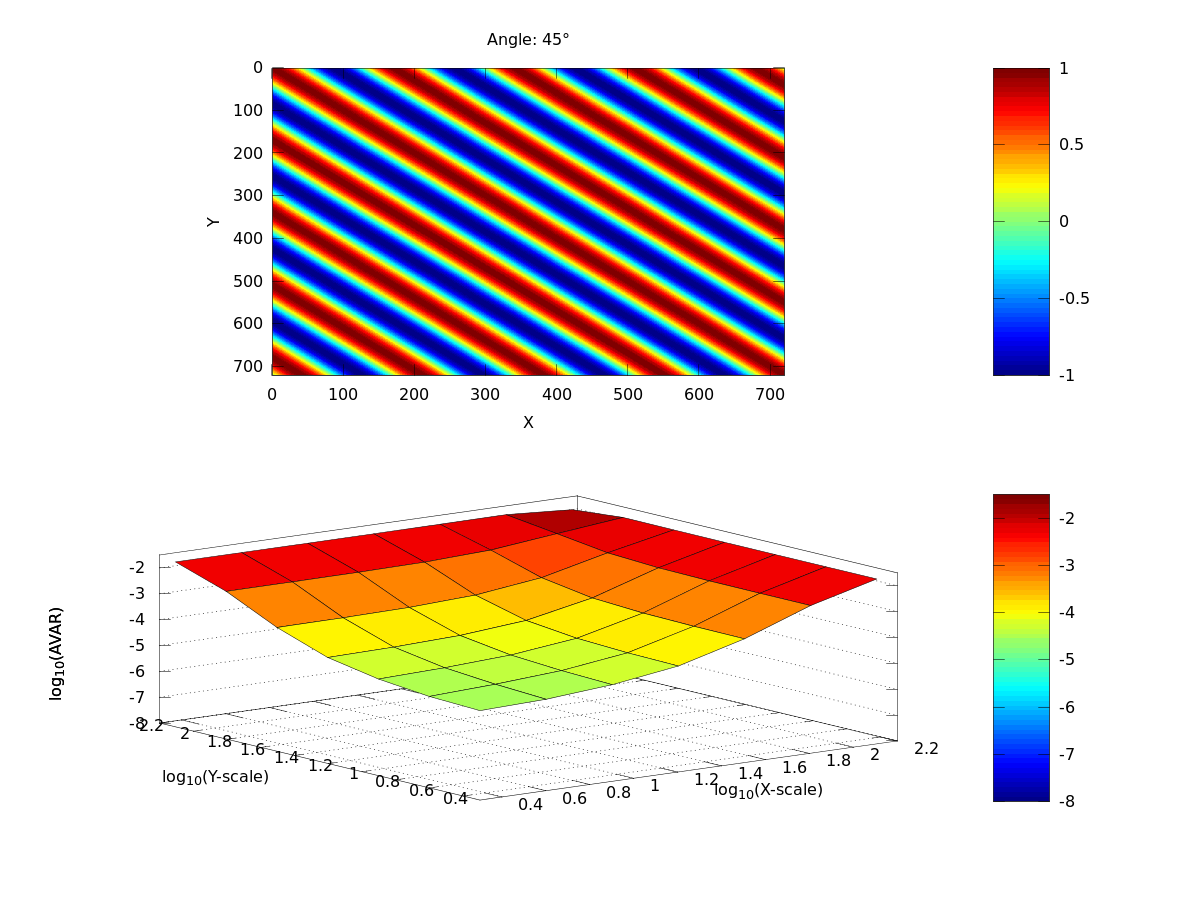}
\caption{Response of Rad-AVAR to a 1D sine with different orientations.\label{fig:radsin1D}}
\end{figure}

	\paragraph{Response to a 1D sine}
The response to a 1D-sine changes slightly with the axis orientation (basically, for a 90$^\circ$ rotation angle, the behavior along the $X$-scale factor is exchanged with the one along the $Y$-scale factor and conversely) but the response is never identically null (see Figure \ref{fig:radsin1D}).

\begin{figure}
\includegraphics[width=8.5cm]{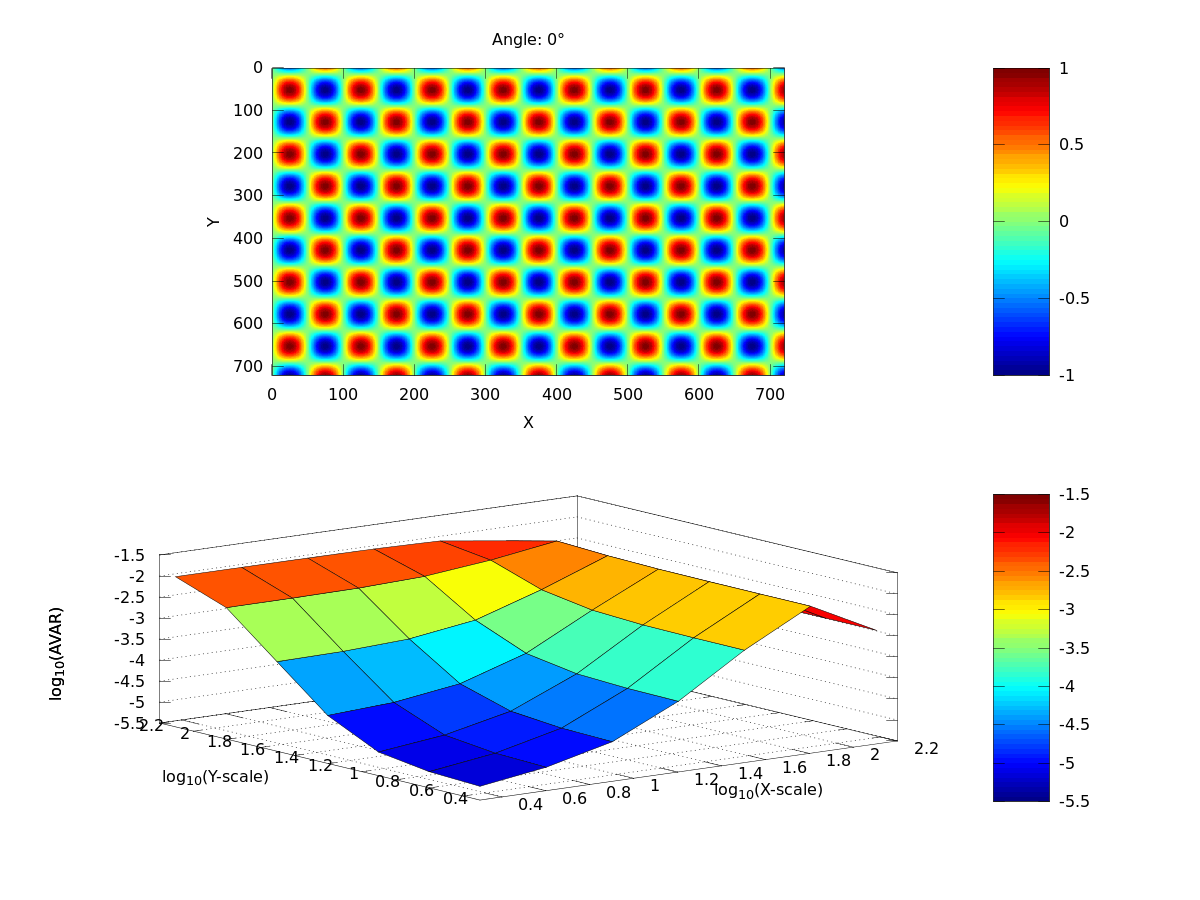} \hfill \includegraphics[width=8.5cm]{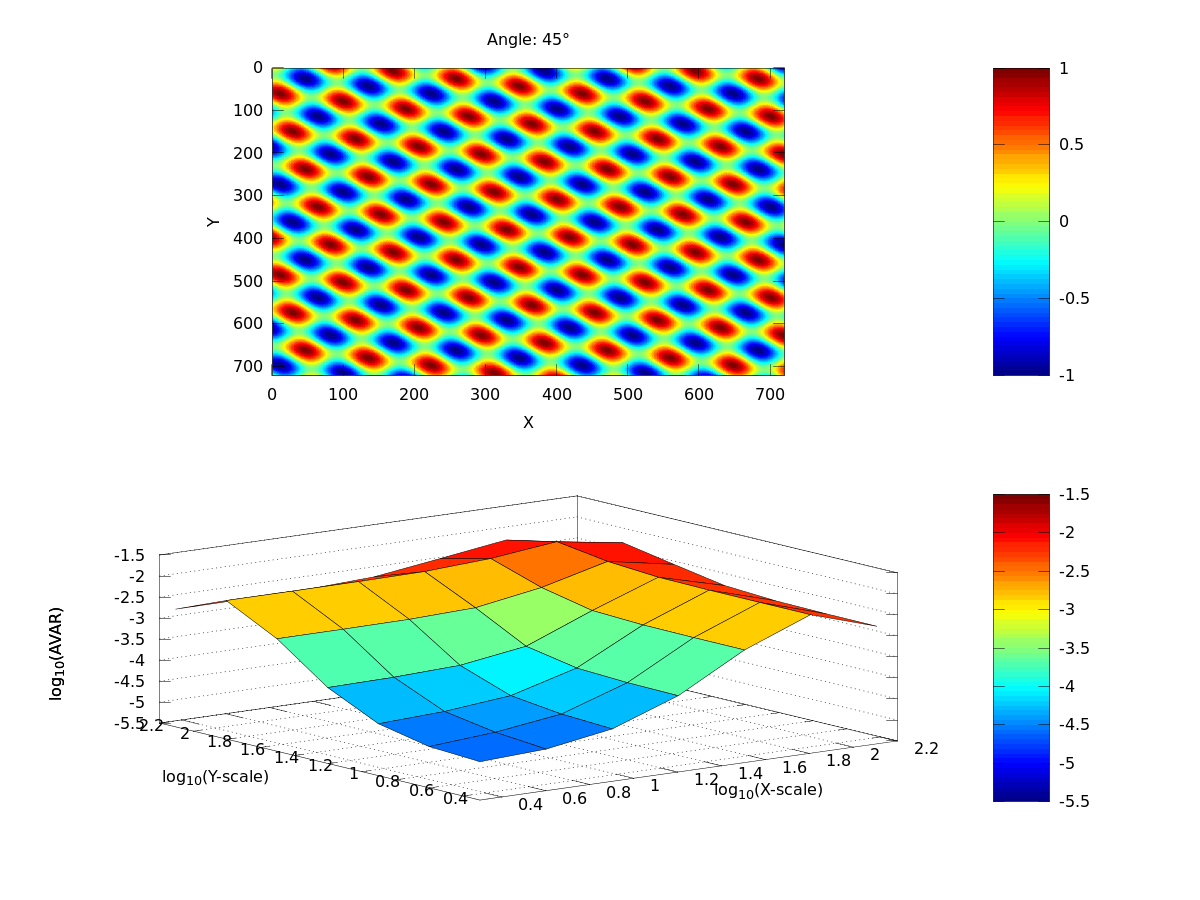}
\caption{Response of Rad-AVAR to a 2D sine with different orientations.\label{fig:radsin2D}}
\end{figure}

	\paragraph{Response to a 2D sine}
Similarly, the response to a 2D-sine changes slightly with the axis orientation but the response is never identically null (see Figure \ref{fig:radsin2D}).

\begin{figure}
\includegraphics[width=8.5cm]{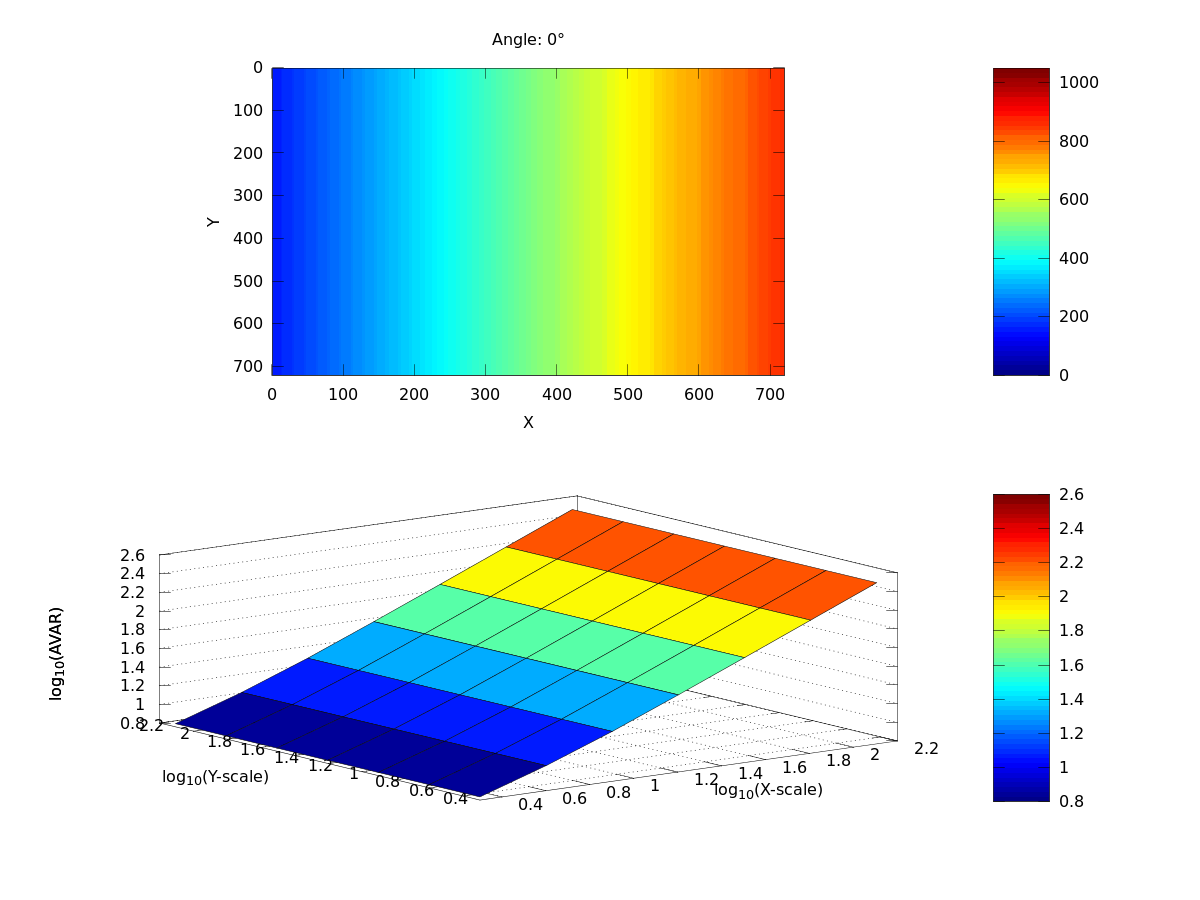} \hfill \includegraphics[width=8.5cm]{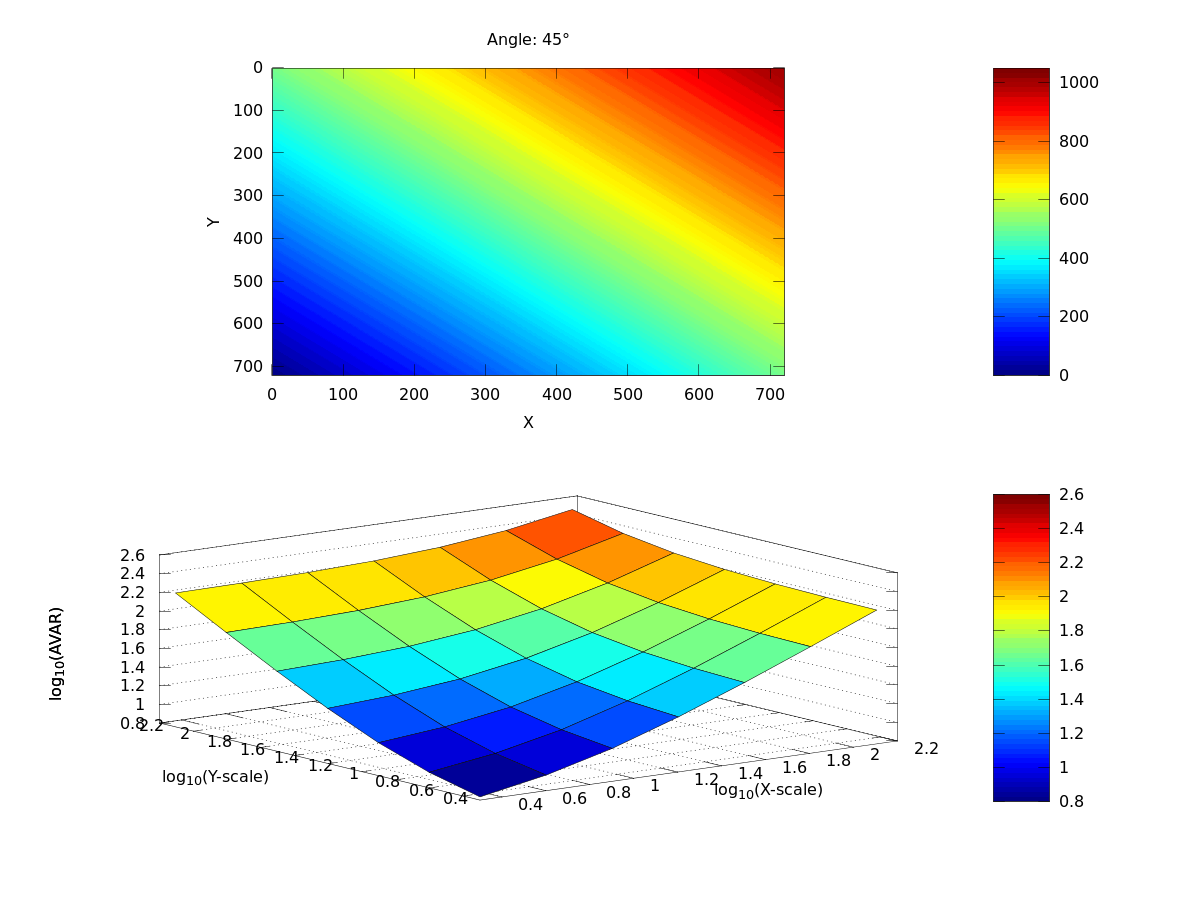}
\caption{Response of Rad-AVAR to a drift with different orientations.\label{fig:raddrift}}
\end{figure}

	\paragraph{Response to a linear drift}
The response passes from a plane increasing proportionally to the $X$-scale factor when the slope of the drift is oriented along the $X$-axis to a plane increasing proportionally to the $Y$-scale factor when the slope is oriented along the $Y$-axis (see Figure \ref{fig:raddrift}). This is then an important property of Rad-AVAR.

	\paragraph{Response to a Gaussian white noise}
The response of Rad-AVAR is a plane decreasing %
with the scale factor as for $XY$-AVAR. Here also, it is almost seen edge-on.

	\paragraph{Response to a Gaussian random walk}
For a random walk, Rad-AVAR gives a quasi flat and horizontal plane: the $Z$-amplitude ranges from -0.9 to -0.7 (from 0.13 to 0.20 on a linear scale) whereas the $X$ and $Y$-scale factors range over 2 decades. The residual fluctuations of the response are due to the difference between the mathematical expectation of Rad-AVAR, which is exactly flat and horizontal, and the estimator of this variance which remains a $\chi^2$ random variable (as for the classical Allan variance). Nevertheless, its fluctuations are very limited and this tool is then entirely able to distinguish between a white noise, a random walk and a geophysical signal.

\begin{figure}
\includegraphics[width=8.5cm]{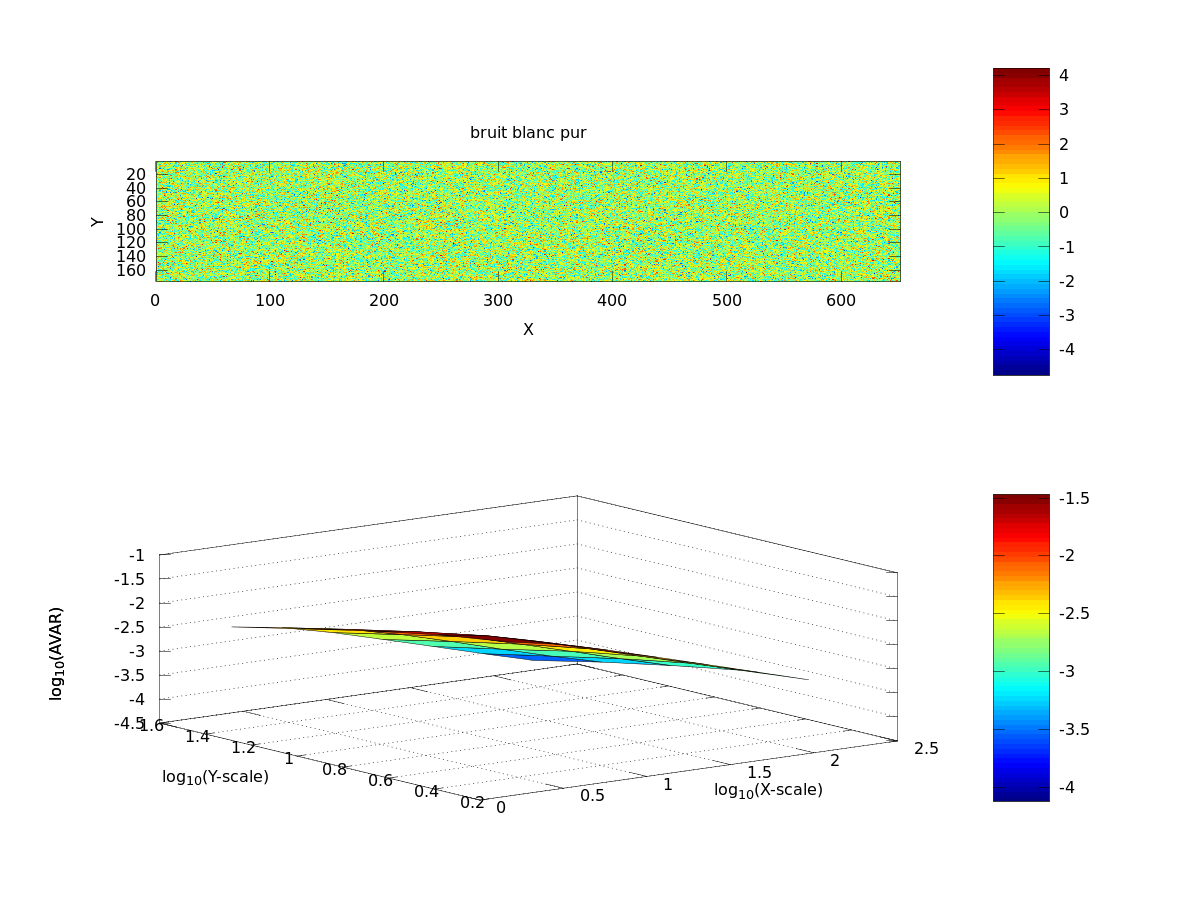} \hfill \includegraphics[width=8.5cm]{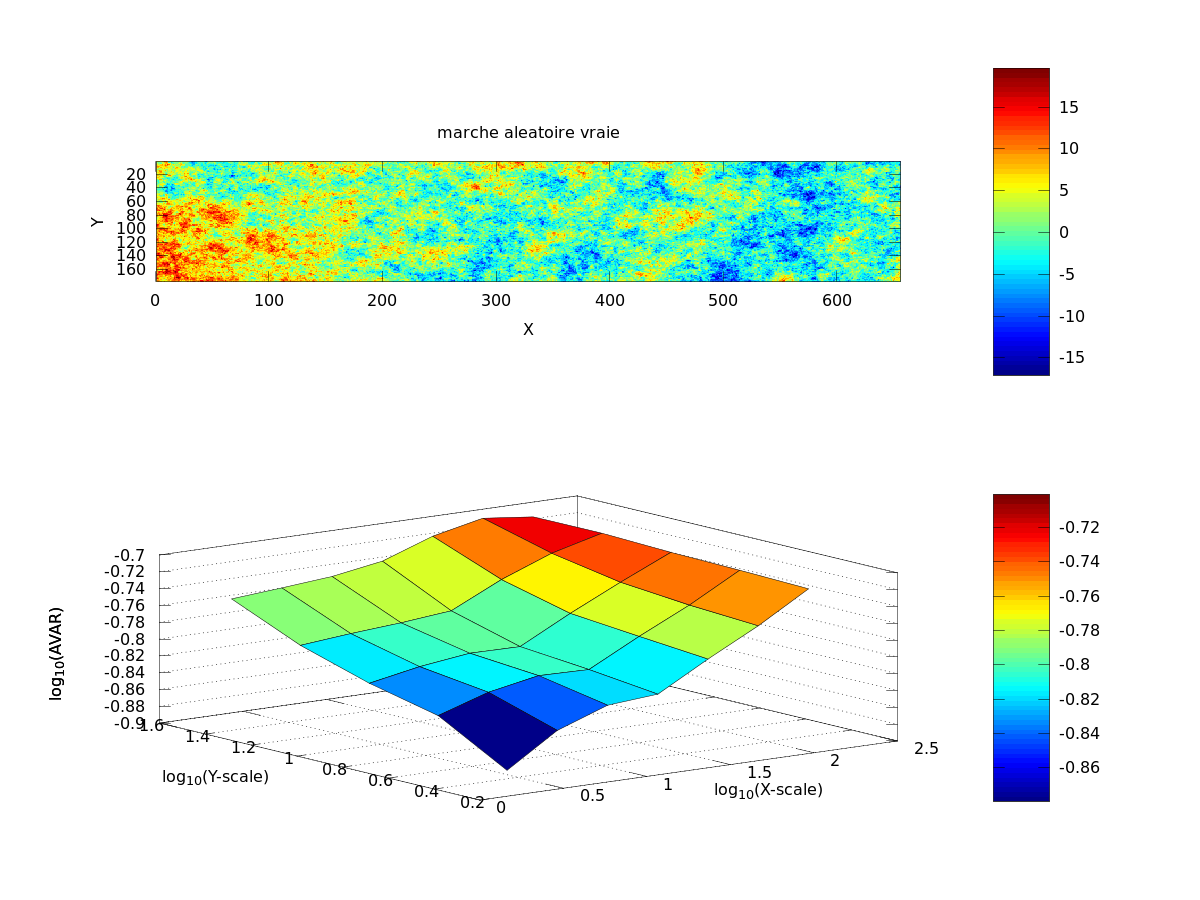}
\caption{Response of Rad-AVAR to white noise (left) and to a random walk (right).\label{fig:radnoise}}
\end{figure}

		\subsubsection{Conclusion about radial AVAR}
Radial AVAR has exactly the expected properties defined in \S \ref{sec:exp_prop}. It is then perfectly suitable for our purpose.
	
\section{Application to InSAR data}
\subsection{Data and method}
\begin{figure}[h]
\includegraphics[width=\linewidth]{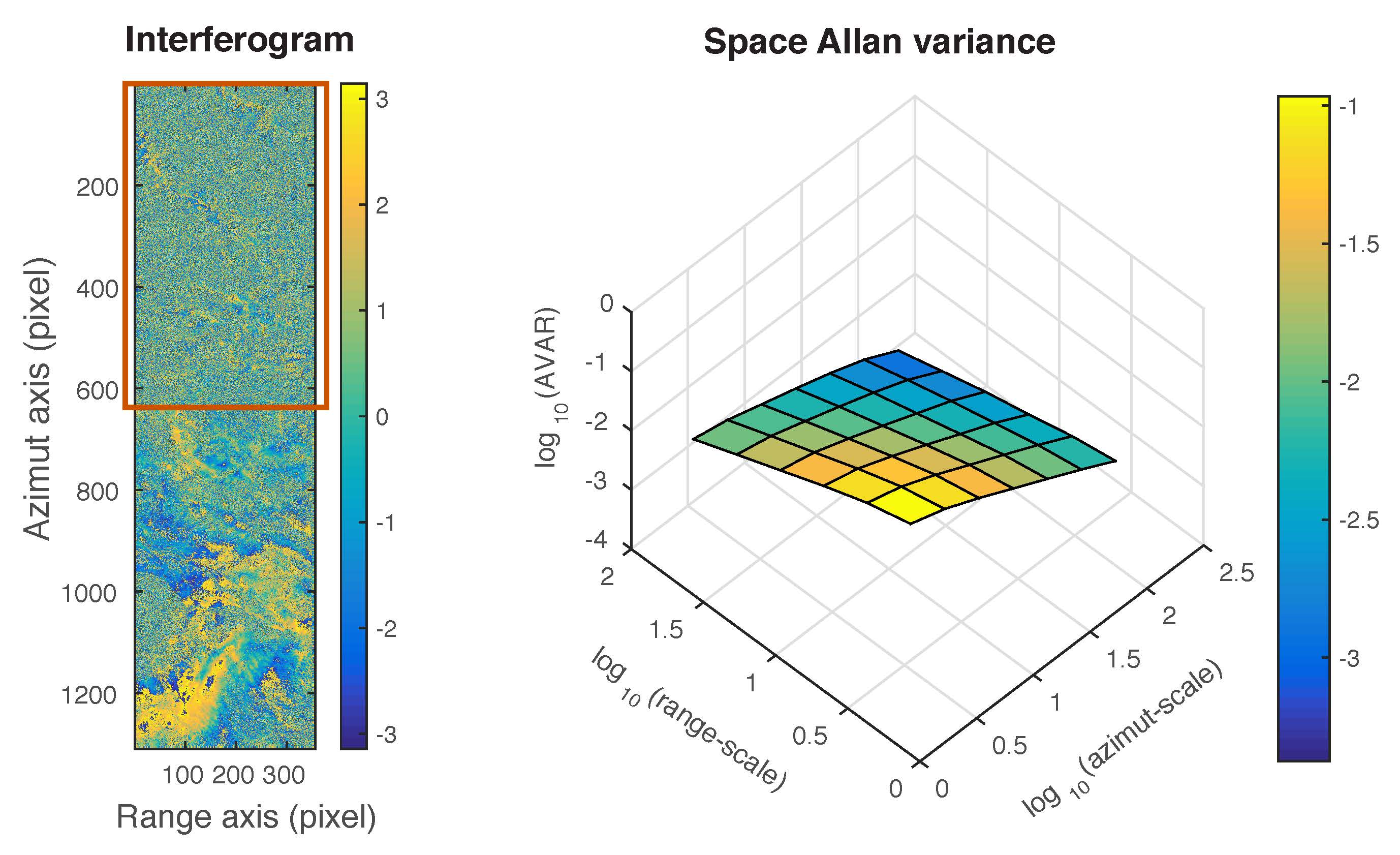}
\caption{\label{fig:white_noise} Right: Interferogram with phase decorrelation (red square). This occurs when spatial baseline is too large between both SAR images or when surface properties (mostly vegetation and erosion) changed between both SAR images acquisition. Left : Space AVAR of the phase values located in the red square. This response corresponds typically to a white noise. }
\end{figure}

We worked with synthetic aperture radar (SAR) images acquired by Envisat, satellite from the European Space Agency (ESA). This dataset has been primarily  used in \cite{cavalie14} to map the surface deformation in eastern Anatolia. 
SAR images are acquired in illuminating the ground with electromagnetic waves (microwave frequency) and collecting the reflected signal. This remote sensing technique uses the round trip travel time and amplitude of the waves to determine physical properties of the targets on the ground that reflected the signal. Two quantities are measured : (i) the amplitude of the reflected signal that corresponds to the energy sent back in the satellite direction and (ii) the phase of the reflected wave. 
Interferograms are  generated in differentiating the phase of two SAR images acquired at different time  \cite{burgmann00}.   Considering that the ground response are identical for both SAR images (i.e, these terms cancel each other out in the phase substraction operation), the observed interferometric phase  $\varphi$ can be written as the sum of four terms  \cite{cavalie08}. We add, here,  one more term to distinguish explicitly between two kinds of atmospheric delays and the other sources of noise :

\begin{equation}
\label{eqn:phs}
\varphi=\varphi_{def}+\varphi_{orb}+\varphi_{strat}+\varphi_{turb}+\varphi_{noise}
\end{equation}

where $\varphi_{def}$ is the phase component related to ground movements, $\varphi_{orb}$ is the residual orbital  phase, $\varphi_{tropo}$ is the stratified atmospheric phase delay correlated with elevation, $\varphi_{turb}$  is the turbulent delays due to the 3-dimensional refractivity distribution of the atmosphere, and  $\varphi_{noise}$ is the remaining term that includes instrument noise, DEM errors, or phase decorrelation. Phase decorrelation occurs when ground surface changes between two SAR acquisitions (vegetation growth, ground erosion) or by geometrical decorrelation when the perpendicular baseline between the two images is too large. When decorrelation happens, wrapped phase takes random values between $-\pi$ and $\pi$, that can be assimilated as white noise. Figure \ref{fig:white_noise} shows phase decorrelation. We obtain a white noise space AVAR signature as obtained in our previous test (Figure \ref{fig:radnoise}).  Such interferograms are then discarded for ground displacement measurements.  

\begin{figure}[h]
\includegraphics[width=\linewidth]{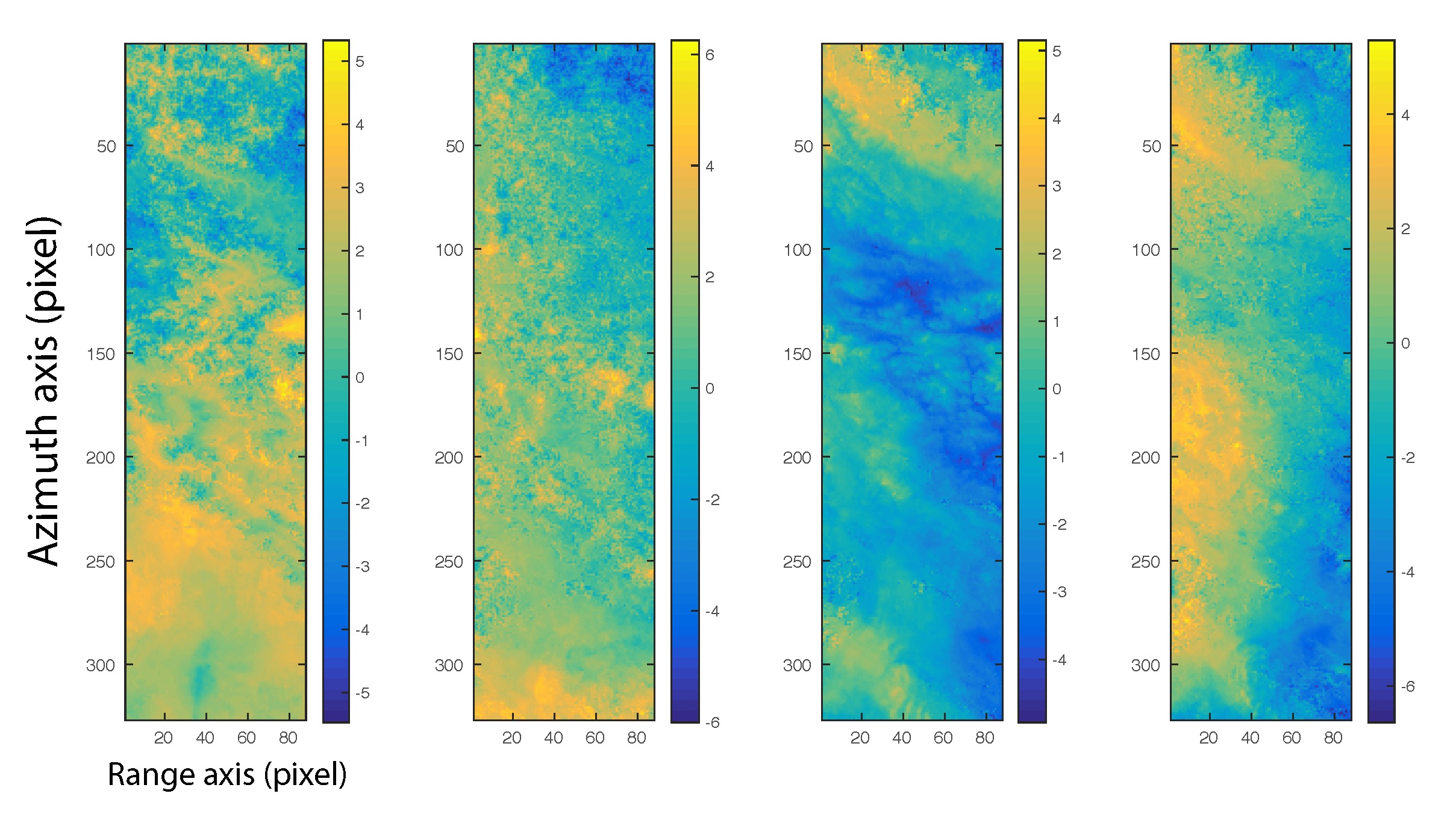}
\caption{\label{fig:ex_igram} Examples of interferograms obtained along track 493. Colorbars indicate phase changes in radians. a) Typical atmospheric patterns due to the turbulent and spatial heterogeneities of the air refractivity. b) On top of the atmospheric delays, higher frequency signal is superimposed due to phase decorrelation. c) Example of miscorrected interferogram where the effect of the stratified atmosphere is still visible (black circle).  d) Interferogram with residual ``orbital ramp'' along the range axis. }
\end{figure}

In our data processing we corrected interferograms for stratified tropospheric delay, $\varphi_{strat}$,  using ERAI, a global atmospheric model \cite{era40}. This correction is a key step to obtaining a successful measurement of tectonic signals by InSAR, although this correction is not always successful and relies on the accuracy of the global atmospheric model. \cite{doin09} shows that ERAI  does not predict properly the tropospheric delay for 10\% of the case. $\varphi_{orb}$ is mostly caused by a drift in the local oscillator onboard the Envisat satellite \cite{marinkovic13}. This drift only causes a ramp in the across-track direction (range) of the interferograms and we therefore made no ramp corrections in the along-track direction (azimuth), to avoid removing tectonic signals related to the motion of the Anatolian and Arabian plates, we estimate the residual orbital signal by an east-west ramp across each interferogram only on their Eurasian part that is considered as stable.   
Others sources of noise can be turbulent atmospheric delays $\varphi_{turb}$, DEM errors or thermal noise, $\varphi_{noise}$. Note that, for the selected interferograms,  $\varphi_{turb}$ is much larger than $\varphi_{noise}$.  Figure \ref{fig:ex_igram} shows single interferograms containing both geophysical signal and noise. As the ground movement is small, noise dominates the scene. Turbulent atmosphere patterns decorrelate within 24 hours \cite{emardson03}. So, %
as SAR images are taken at different dates (shortest repetitive acquisition is 35 days), interferograms that do not share common images, will exhibit different noise signal patterns (Figure \ref{fig:ex_igram}). 

\begin{figure}[h]
\includegraphics[width=\linewidth]{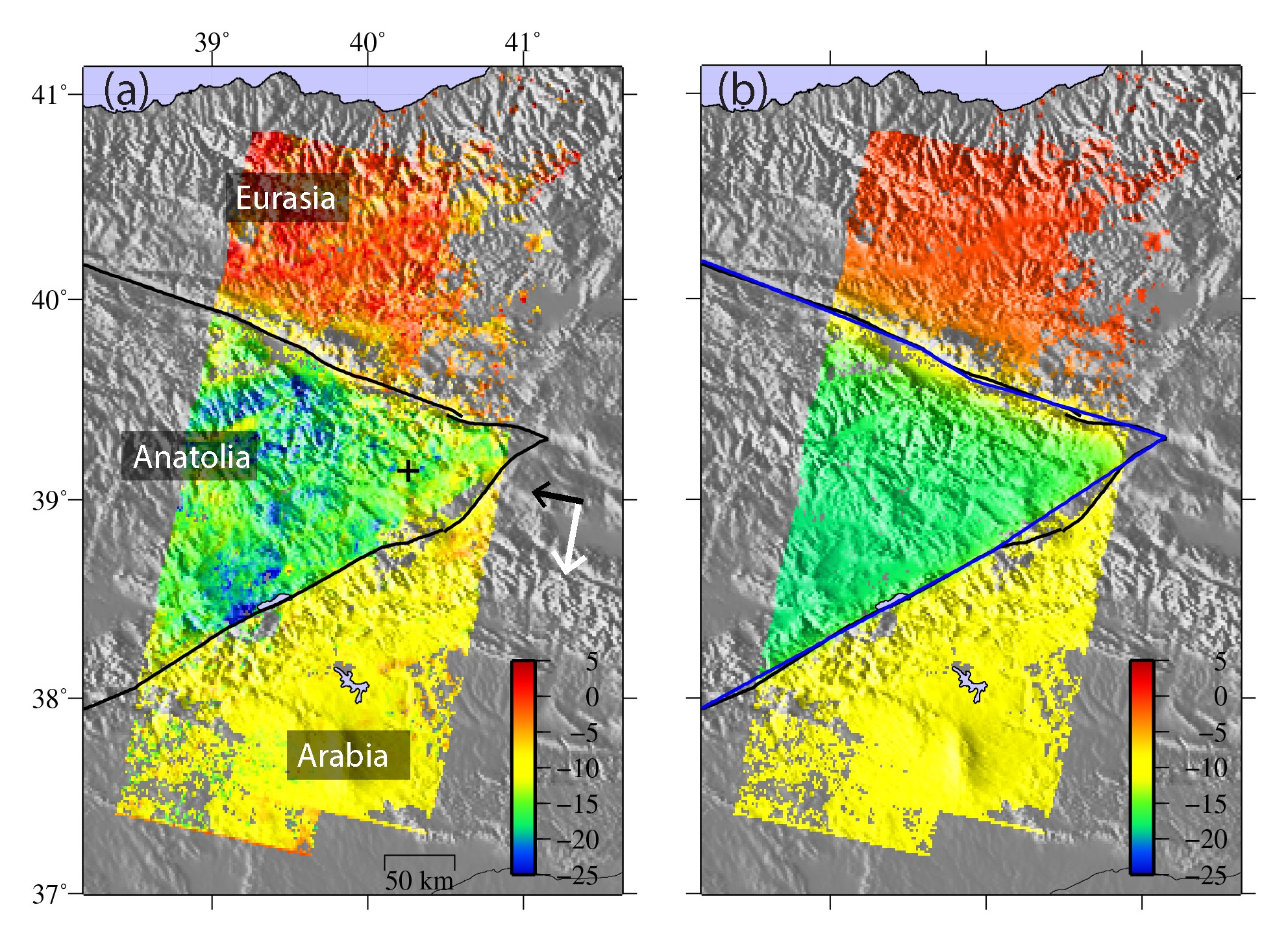}
\caption{\label{fig:geo}Observed and modeled ground velocities. (a) Observed  horizontal velocities from InSAR time series analysis in mm/yr parallel to the satellite line of sight (LOS) direction (black and white arrows show the LOS and flight directions, respectively) and in an Eurasia-fixed reference
frame. Velocities are shown positive in the east-southeast direction (103 N). (b) Elastic back slip model modeling. The black cross in (a) points the location of the time series of Figure \ref{fig:disp}.\label{fig:geo_X}}
\end{figure}

To derive ground %
displacement rates from noisy interferograms,  \cite{cavalie14} used small baseline time series analysis \cite{doin12}. This method uses interferograms, computed with the same image dataset, to retrieve phase increment between two images acquired successively in time.  As said previously, most of the remaining interferogram noise is high frequency in time. A smoothening operator is then applied to limit phase variations due to turbulent atmospheric delays. This step allows to estimate the surface displacement history of each coherent pixel from the set of small-baseline interferograms.   
For tectonic application, \cite{cavalie14} computed the linear component of the time series result for each pixel, to derive the steady surface displacement velocities. This result can be explained by simple elastic model where plate borders are represented by buried dislocations, creeping at a rate allowing to accommodate relative plate motions (i.e, here Eurasia, Anatolia and Arabia). Figure \ref{fig:geo} shows observed (by InSAR) and modeled velocities maps. Note that maps are shown here in the geographic referential and the phase has been converted in centimeters (for Envisat satellite, 2$\pi=2.8$ cm). For simplicity reasons, we will work with radar coordinates in the following and phase amplitudes will be in radian. 

We tested the Allan Variance on various InSAR items.  First on synthetic tests, to see how the 2D-AVAR reacts, %
 and then on the real InSAR data presented above. 

\subsection{InSAR signal modeling} 

\subsubsection{Tectonic signal modeling}

\begin{figure}[h]
\includegraphics[width=\linewidth]{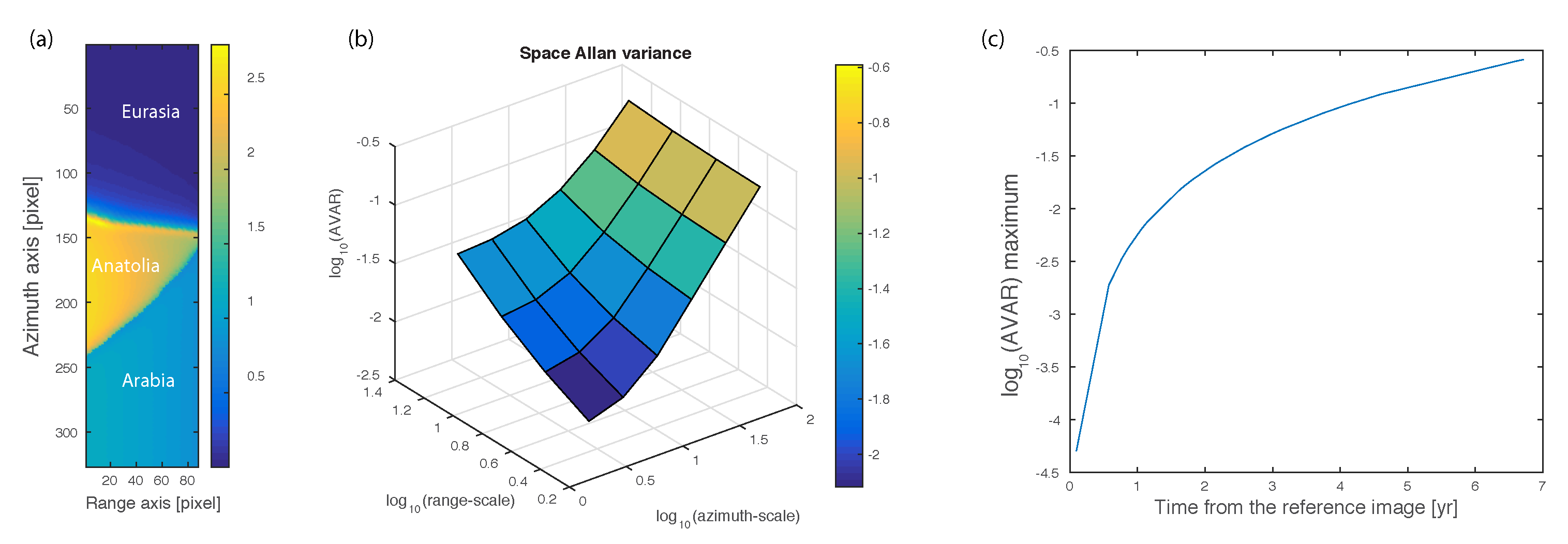}
\caption{\label{fig:avar_model1} (a) Modelled tectonic signal in radar coordinates corresponding to the line of sight (LOS) ground displacement. (b) Space Allan variance computed from (a). (c) Evolution of log$_10$(AVAR) maximum value for each SAR acquisition date.}
\end{figure}

To have an idea of how the 2D-AVAR of the geophysical signal evolves with time, we modeled the tectonic signal in radar coordinates for each SAR image date. The spatial AVAR is then computed at each time step. The first SAR acquisition of the time series is our reference image. There is thus no deformation. We will then consider the deformation between the date of each other images and this reference. We assume the ground displacement to be linear in time. So, knowing the plate velocities, it is then possible to compute the expected geophysical signal between each SAR image and the reference image (Figure \ref{fig:avar_model1}a). As the tectonic signal has a large spatial wavelength content and increases linearly with time, we expect the AVAR to increase with time and as the scale factor increases.
	
Tectonic signal is mostly a large wavelength signal. Applying the spatial AVAR on the modeled signal, we see that the variance increases both in azimuth and range axis. However, du to the geometry and movement of the plates, AVAR reaches larger values along the azimuth axis (Figure \ref{fig:avar_model1}b).   As the signal grows in time, the shape and AVAR range do not change, %
but the maximum AVAR value, found for large scale factor, increases (Figure \ref{fig:avar_model1}c).

\subsubsection{Atmospheric noise modeling} 
Based on Tatarski's work \cite{tatarski61}, notably, Hanssen proposes a new atmospheric model for radar interferometry considering three regimes of atmospheric delays as observed by InSAR \cite{hanssen2002}. Each of these regimes corresponds to a particular scale factor and is characterized by an exponent of the power spectrum $\beta$:
\begin{description}
	\item[I] scale-factor $> 2$ km: $P_\phi(k) \propto k^{-5/3}$
	\item[II] $0.5 <$ scale-factor $< 2$ km: $P_\phi(k) \propto k^{-8/3}$
	\item[III] $0.01 <$ scale-factor $< 0.5$ km: $P_\phi(k) \propto k^{-2/3}$
\end{description}
where $k$ is the wave number and $P_\phi(k)$ the power spectrum. This model assumes isotropy and no ionospheric influences, that is a verified for C-band SAR data as used in this study. However, the model does not take into account topography related delays, $\varphi_{tropo}$, what might be a problem if this term is not properly removed.

We can, thus, model the atmospheric noise by the following dependence on the wave number:
$$
P_\phi(k)=\frac{k}{k+k_{12}}\left(h k^{-8/3}+h_{3}k^{-2/3}\right)
$$
where 
\begin{itemize}
	\item $k_{12}$ is the cut-off wave number between regimes I and II: $k_{12}=0.5$ km$^{-1}$
	\item $h$ is a free parameter denoting the general level of atmospheric noise
	\item $h_3$ is the level of the regime III which value is fixed by the previous parameter: since the limit between regimes II and III corresponds to scale-factor of $0.5$ km, i.e. $k_{23}=2$ km$^{-1}$, it yields
$$
h k_{23}^{-8/3}=h_{3}k_{23}^{-2/3} \quad \Rightarrow \quad h_3=h k_{23}^{-2}=h/4.
$$
\end{itemize}
The atmospheric noise model is then:
\begin{equation}
P_\phi(k)=\frac{h k\left(k^{-8/3}+k^{-2/3}/4\right)}{k+0.5}\label{eq:hanssensmodel}
\end{equation}
where $k$ is expressed in km$^{-1}$.

In our case, a pixel corresponds to 640 m and an image contains typically 600 pixels. Thus, $k$ ranges between $0.0025$ and $0.8$ km$^{-1}$ (Nyquist frequency).

\begin{figure}
\includegraphics[width=\linewidth]{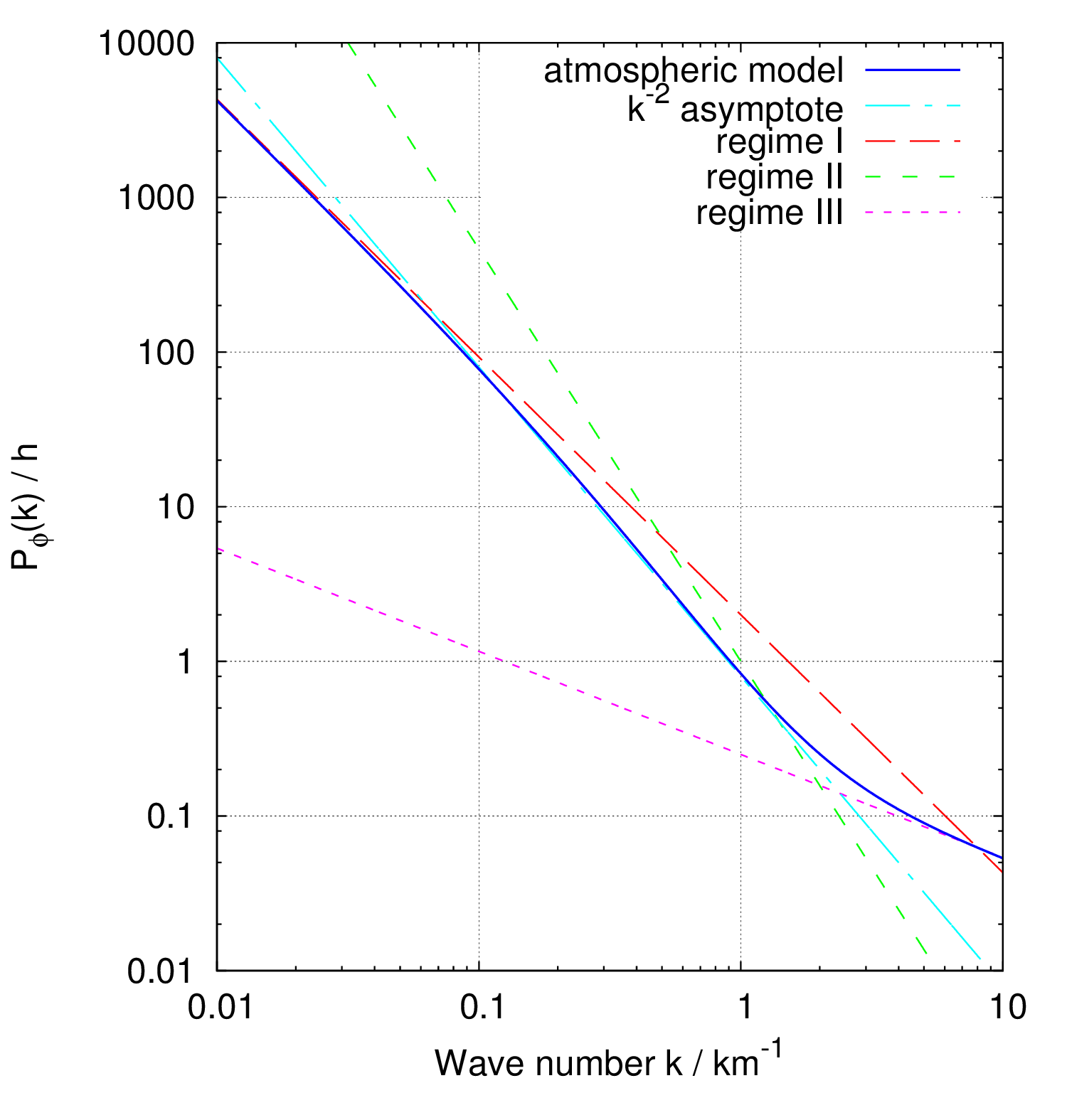}
\caption{Model of atmospheric noise. The three regimes are visible: 
regime I at left (for $k<0.05$ km$^{-1}$, dashed red asymptote), regime II near the center ($k\approx 0.5$ km$^{-1}$, dashed green asymptote), regime III at right (for $k>5$ km$^{-1}$, dashed purple asymptote). However, the atmospheric noise model remains relatively close to the $k^{-2}$ asymptote (dash-dotted cyan line). \label{fig:atmonoise}}
\end{figure}

Figure \ref{fig:atmonoise} shows the Hanssen's model. It is very interesting to notice that this model remains close to a $k^{-2}$ asymptote within the $k$-range which corresponds to our images. It means that the response of the atmospheric noise to space AVAR should be very similar to a plan almost flat and horizontal which confirms our intuition of \S \ref{sec:exp_prop}.

\begin{figure}[h]
\includegraphics[width=\linewidth]{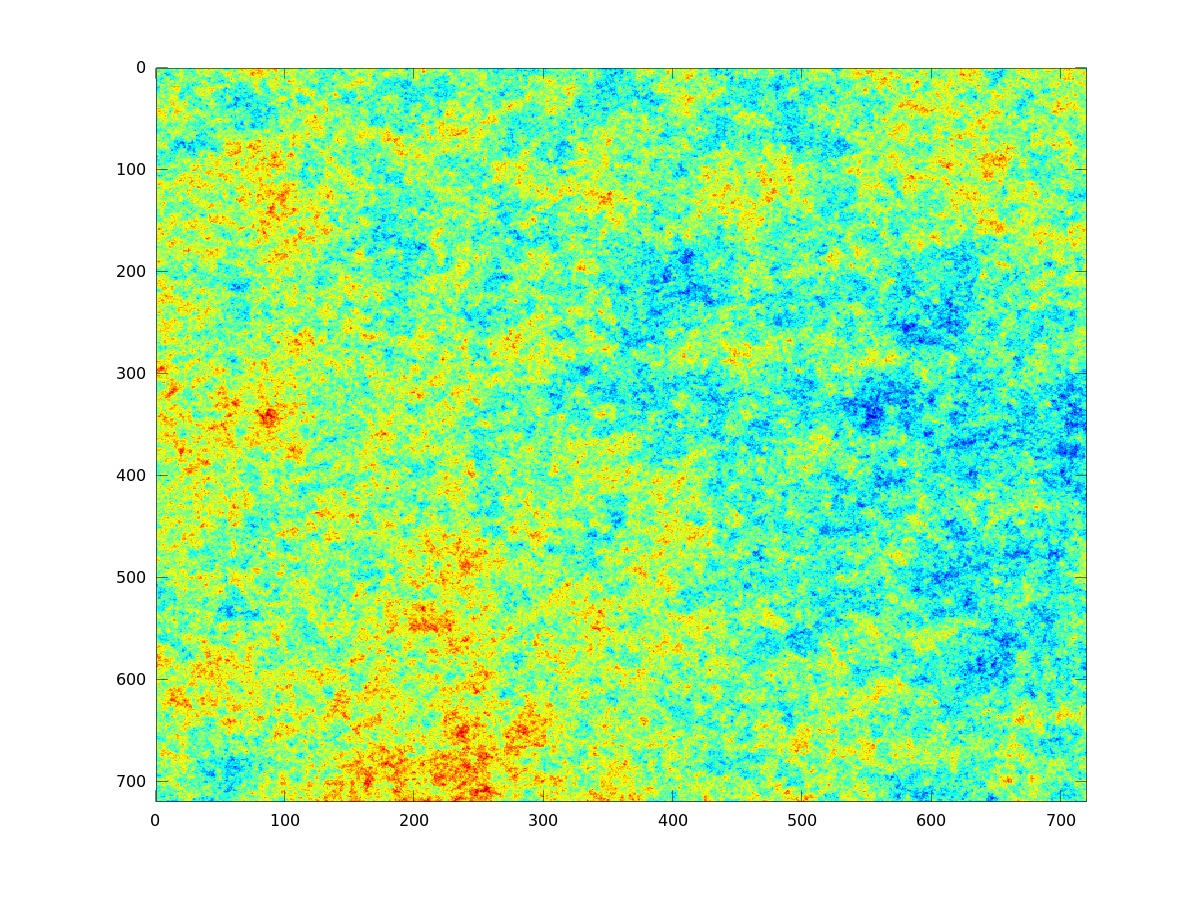}
\caption{Simulation of an atmospheric noise according to the Hanssen's model described by relationship (\ref{eq:hanssensmodel}).\label{fig:exatmonoise}}
\end{figure}

\begin{figure}[h]
\includegraphics[width=\linewidth]{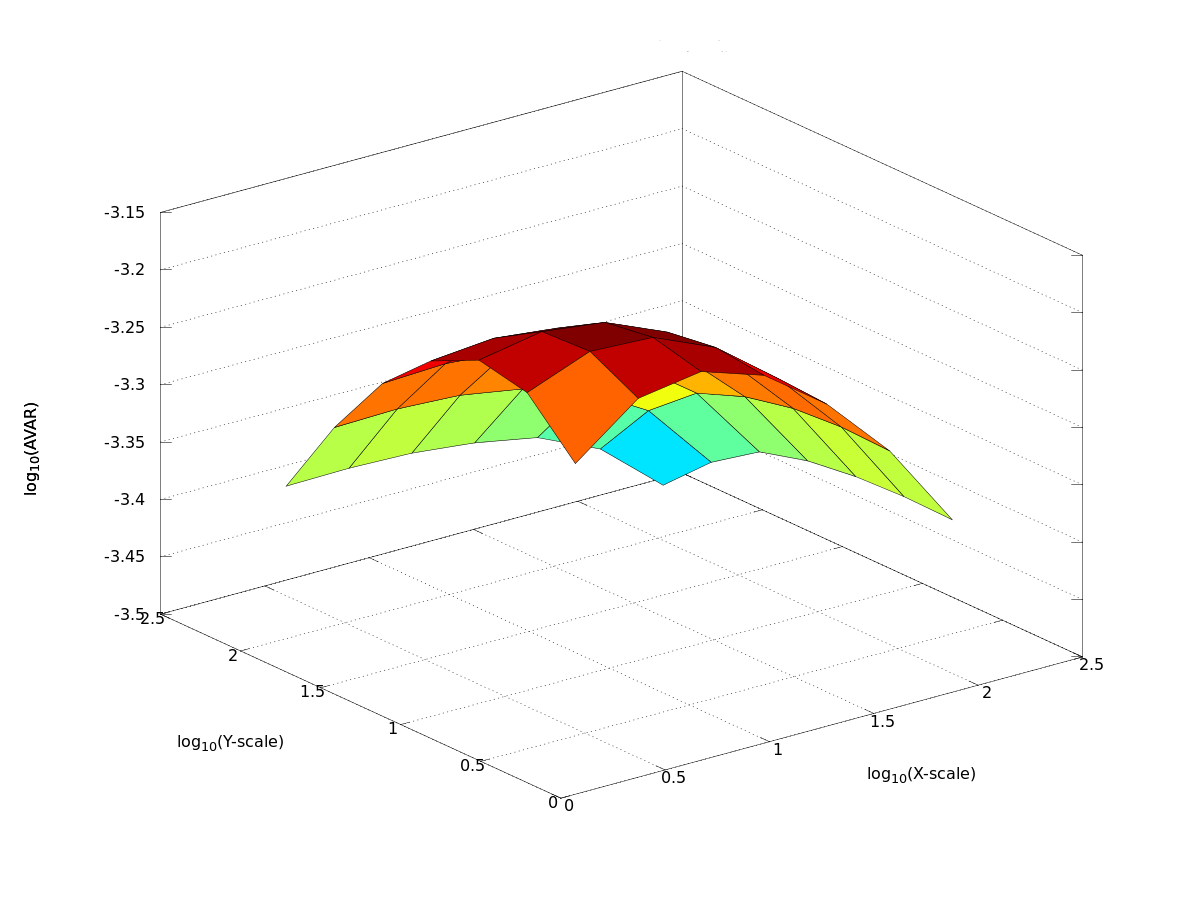}
\caption{Response of space AVAR for an atmospheric noise. This response is the average of the response of AVAR for 25 uncorrelated simulated atmospheric noises (Figure \ref{fig:exatmonoise} is on these). It must be noticed that $z$-scale ranges from $10^{-3.5}$ to $10^{-3.15}$, i.e. from $3\cdot 10^{-4}$ to $7\cdot 10^{-4}$, which proves that this response is nearly flat.\label{fig:atmoresp}}
\end{figure}

This model may be used for simulating images of atmospheric noise. Such an image is displayed in Figure \ref{fig:exatmonoise}. The space Allan variance of this sort of images may thus be computed. Figure \ref{fig:atmoresp} exhibits the average of the response of AVAR for 25 uncorrelated simulated atmospheric noises like Figure \ref{fig:exatmonoise}. Averaging 25 responses cancels out the deviation of each individual response from the theoretical variance of space AVAR. Thus, Figure \ref{fig:atmoresp} is representative of the response of space AVAR for an atmospheric noise following the Hanssen's model.

\begin{figure}[h]
\includegraphics[width=\linewidth]{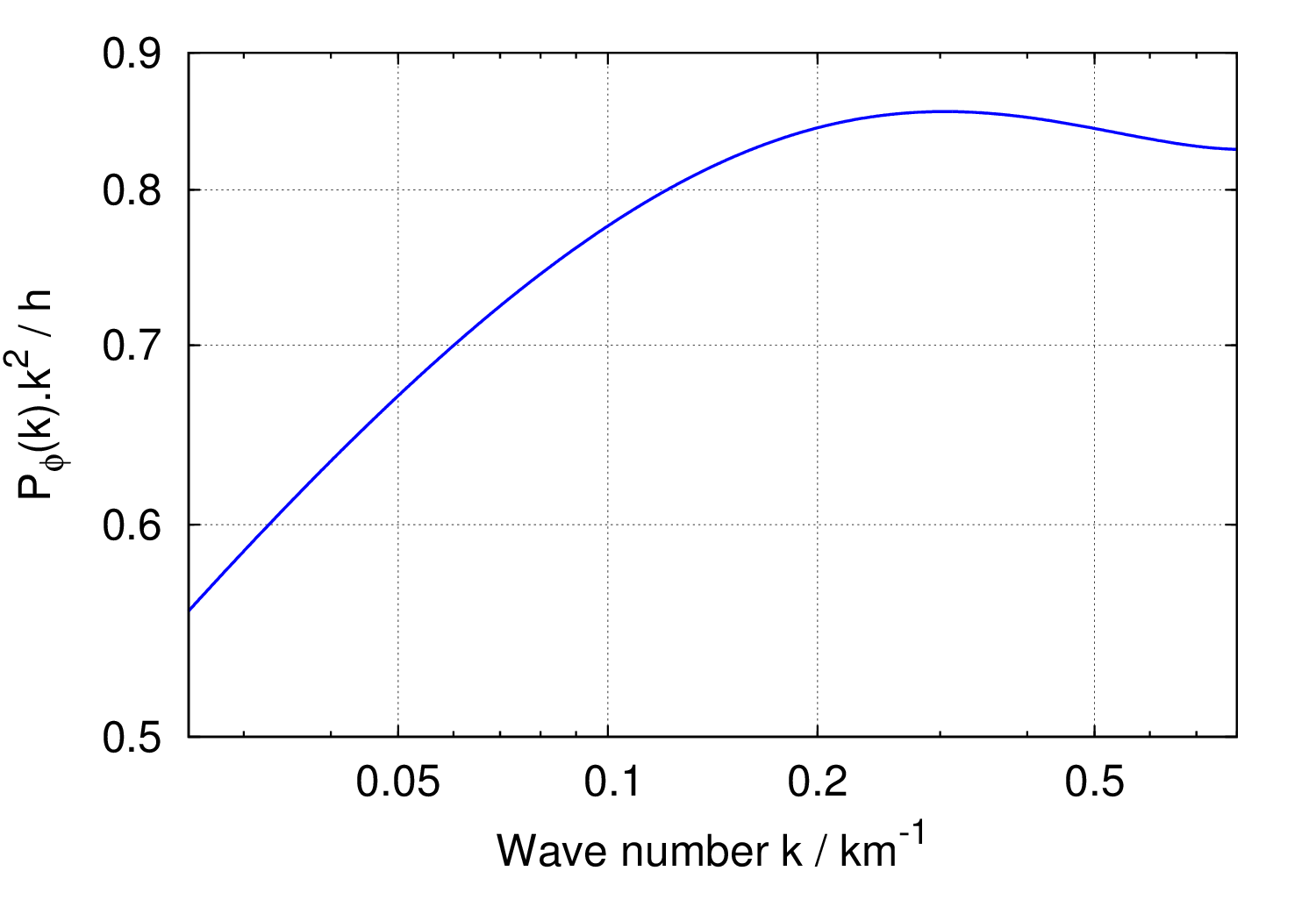}
\caption{Power spectrum of the atmospheric noise model multiplied by $k^2$ within the observable range. The $k^2$ factor allows us to clearly identify that the power spectrum follows a $k^{-2}$ power-law (horizontal asymptote) around $k=0.3$, a power-law with an exponent $>-2$ for lower $k$, and a power-law with an exponent $<-2$ for higher $k$.\label{fig:modk2}}  
\end{figure}

As expected, taking into account the very thin range of Figure \ref{fig:atmoresp} $z$-scale, we can see that the response of space AVAR is almost horizontal and flat. More precisely, it looks like a cupola: horizontal near the center and decreasing on the edges. For going further, we must analyze Figure \ref{fig:modk2} which shows the power spectrum of the atmospheric noise multiplied by $k^2$, in order to identify the areas where the slope follows a $k^{-2}$ asymptote rendered as an horizontal curve. Indeed, this figure shows that the atmospheric power spectrum follows exactly a $k^{-2}$ asymptote around $k=0.3$ km$^{-1}$, yielding a flat response of space AVAR for scale factors around $5$ ($0.7$ on log$10$-scale). For larger $k$, the curve of Figure \ref{fig:modk2} exhibits a decreasing slope: this means that the power-law of the power spectrum is lower than $-2$ yielding a positive slope for space AVAR at small scale-factors. On the other hand, for $k$ smaller than $0.3$, the curve of Figure \ref{fig:modk2} exhibits an increasing slope: the power-law of the power spectrum is then higher than $-2$ and space AVAR decreases as the scale factor increases.

We can now compare this theoretical model with single interferograms  acquired over the area presented on Figure \ref{fig:geo}. For that purpose, we chose interferograms %
with small temporal baseline (i.e small timespan between the two SAR acquisitions) in order that the tectonic signal is negligible with respect to the atmospheric delays. Space AVAR would thus reflect the atmospheric structure. Figure \ref{fig:noise_struct} shows that typical space AVAR pattern for a corrected interferogram (i.e $\varphi_{strat}$ and $\varphi_{orbit}$ have been removed) differs from Hanssen's model at large scale factors. Space AVAR response of the digital elevation model (DEM) also displays a quasi-plane increasing with scale factors. So even, if DEM related components have been removed, residual of this component might still be strong enough to contaminate the space AVAR of the APS response.  Space AVAR shows thus that in areas where strong  elevation variations take place, like in Turkey, the influence of the topography cannot be neglected  to model the atmosphere for InSAR. 
However, this effect must be relativized since the magnitude of this increasing of space AVAR for high scale factors is limited to half a decade (from $10^{-4.5}$ to $10^{-4}$, i.e from $3\cdot 10^{-5}$ to $10^{-4}$).

An empirical way has also been proposed to model the atmospheric topology for InSAR \cite{sudhaus09}. This method estimates directly the variances and autocovariances of interferograms. For this purpose, semi-variograms $\gamma(h)$ is used  to estimate the InSAR
variances and covariograms $C(h)$ is used to estimate the spatial correlation in the data \cite{chiles99}. They are defined as followed : 

\begin{equation}
\gamma(h)=\frac{1}{N} \sum_{i,j \atop d(i,j)=h}|\varphi(i)-\varphi(j)|
\end{equation}

where $\varphi(i)$ and $\varphi(j)$ are the interferometric phases of pixels $i$ and $j$, respectively, $d(i, j)$ is the distance beween $i$ and $j$, and $N$ is the
number of pixel pairs such that $d(i, j)=r$. Similarly, one can define the covariogram as:

\begin{equation}
C(h)=\frac{1}{N} \sum_{i,j \atop d(i,j)=h} \varphi(i) \cdot \varphi(j)
\end{equation}

\begin{figure}[h]
\includegraphics[width=\linewidth]{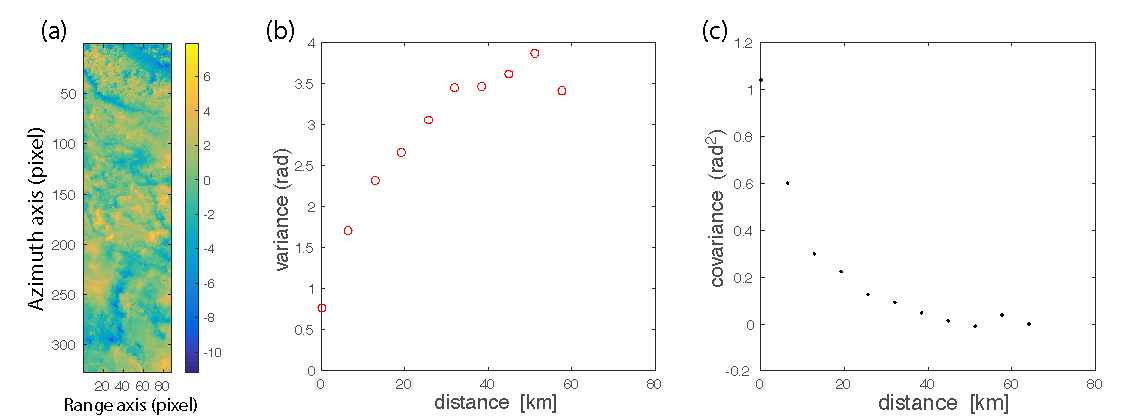}
\caption{\label{fig:var_covar} (a) Interferogram computed with images acquired with a 35-day timespan. (b) Semi-variogram, $\gamma(h)$ and (c) covariogram, $C(h)$, computed for this interferogram. }
\end{figure}

From the semi-variograms, one can estimate the variance of the interferograms when the function flattens (Figure \ref{fig:var_covar}). Covariogram indicates the correlation length in the data and how fast it decorrelates. In case of white noise (i.e no correlation), covariogram function would be constant-zero function and semi-variogram would be a constant function at the variance value. 
Covariagram function can be estimated by a an exponential decay $b\cdot \exp(-h/a)$.  The exponential decay function is then used to generate random, similar spectral, noise (Figure \ref{fig:noise_struct}).

\begin{figure}[h]
\includegraphics[width=\linewidth]{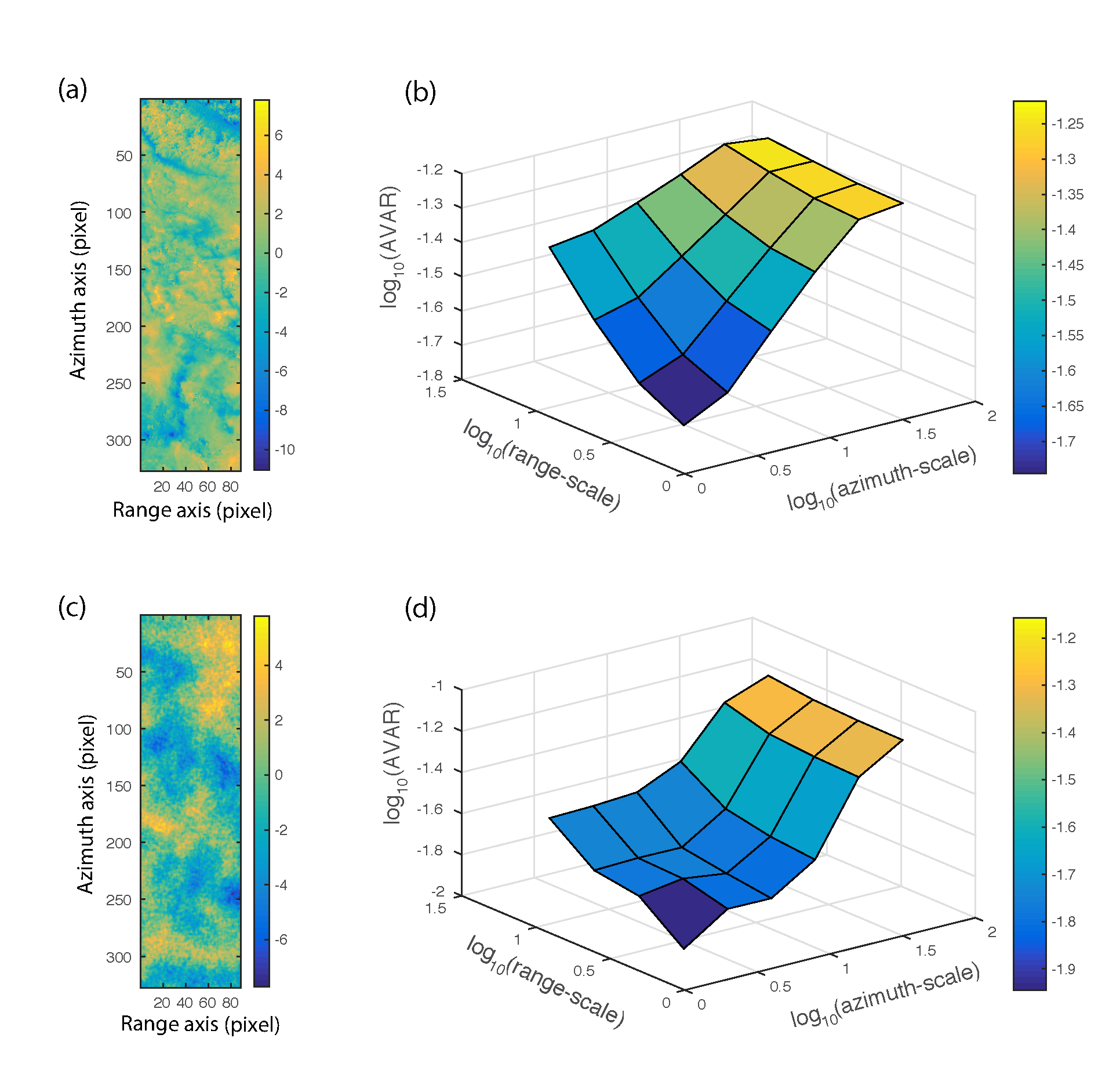}
\caption{\label{fig:noise_struct} (a) Same interferogram as in Figure \ref{fig:var_covar}a (b) corresponding space AVAR diagram. (c) Modeled atmospheric noise based on the interferogram covariogram and (d) corresponding space AVAR diagram. }
\end{figure}

\subsection{Real data}

\begin{figure}[h]
\includegraphics[width=\linewidth]{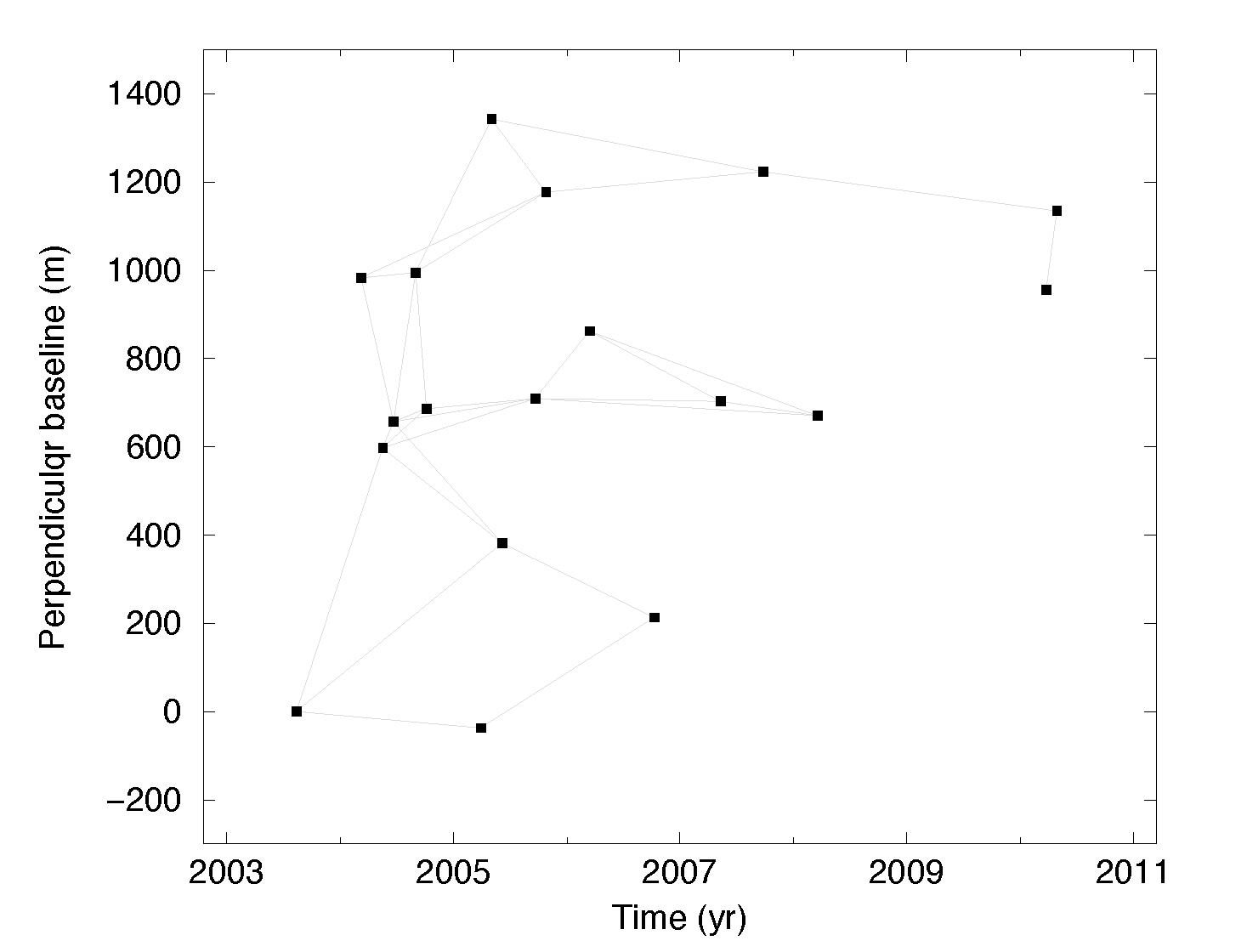}
\caption{\label{fig:baseline} Diagram of Envisat images for descending track 493. Relative perpendicular baselines are plotted as a function of acquisition dates. The
18 SAR images are combined into 32 interferograms (grey segments).}
\end{figure}

We now applied the 2D-AVAR on data acquired from Envisat satellite over an area covering the eastern part of the Anatolian plate. 
 We combined 18 SAR images into 32 interferograms. Figure \ref{fig:baseline} shows image acquisitions date as a function of their relative position (perpendicular baseline). Note that to keep a good phase coherence over the interferogram, perpendicular and temporal baseline between two images is limited. Based on this criteria, we find the best network that connect each image. Each interferogram is corrected for $\varphi_{orb}$ and $\varphi_{strat}$ \cite{cavalie14}. We computed for each interferogram the spatial AVAR before and after the correction. We observe that the correction step clearly decreases the mean AVAR, confirming that large wavelength have been reduced. The next step is to inverse the set of interferograms to establish a time series of the phase evolution. 

\subsection{Data Inversion}
\subsubsection{Inversion without smoothing}
Corrected interferograms are then inverted to solve for the incremental deformation between two successive  images using a least squares inversion method \cite{menke_book}. For each pixel, treated independently 
from its neighbours, we solve: 
\begin{equation}
\label{eqn:dgm}
\textbf{d}=\textbf{G}. \textbf{m}
\end{equation}
where \textbf{d} includes $N$ interferograms observations, \textbf{m} corresponds to  $M$-1 incremental displacements  between the $M$ time steps,  and \textbf{G} is a $N$  by $M-1$ matrix  containing zeros and ones, based on the stating that the interferometric phase, 
$\varphi_{ij}$,  is the sum of successive phase increments between image $i$ and image $j$ : 
$\varphi_{ij}=\sum_{k=i}^{j-1}m_k$. Images are ordered by dates.\\
 
The inversion uses the redundant information of the 32 interferograms to reduce some errors due to interferogram formation.

\subsubsection{Inversion with smoothing}
 
To decrease atmospheric delay or APS, noted $\varphi_{noise}$ in equation \ref{eqn:phs}, we introduce temporal smoothing in the inversion as another constraint, by minimizing the curvature of the inverted phase temporal evolution \cite{schmidt03}.  The system becomes :
\begin{equation}
\label{eqn:inversion}
  \left(
       \begin{array}{c}
        \textbf{d}'  \\
        0  \\
        \end{array}
        \right)     =\left(
       \begin{array}{c}
        \textbf{G}' \\
        \gamma^2\omega_i\frac{d}{dt^2}  \\
        \end{array}
        \right).\left(
        \begin{array}{c}
         \textbf{m} \\
         \end{array}
         \right)
\end{equation}

\textbf{d},  \textbf{G}, and \textbf{b} are weighted by the matrix \textbf{W} 
($\textbf{d}'=\textbf{W}\textbf{d},\ \textbf{G}'=\textbf{W}\textbf{G}$,  and  
$\textbf{b}'=\textbf{W}\textbf{b}$).  
$\gamma$ is the smoothing coefficient introduced to ponderate the minimum curvature constraints  $\frac{dm}{dt^2}$ (where $t$ is time).  $\left.\frac{dm}{dt^2}\right|_i$ is evaluated with a 5 points 
finite difference scheme centered on each acquisition date $i$.  The weight, \textbf{W}, applied on each interferogram is the product of two terms  :  
(1) the first term equalizes the weight of all images in the inversion, 
(2) the second term characterizes the interferograms ``quality'', $q_{ij}$, defined as 
the inverse sum of both images $i$ and $j$ APS amplitudes. \\

\subsubsection{Times series results and spatial AVAR}

\begin{figure}[h]
\includegraphics[width=\linewidth]{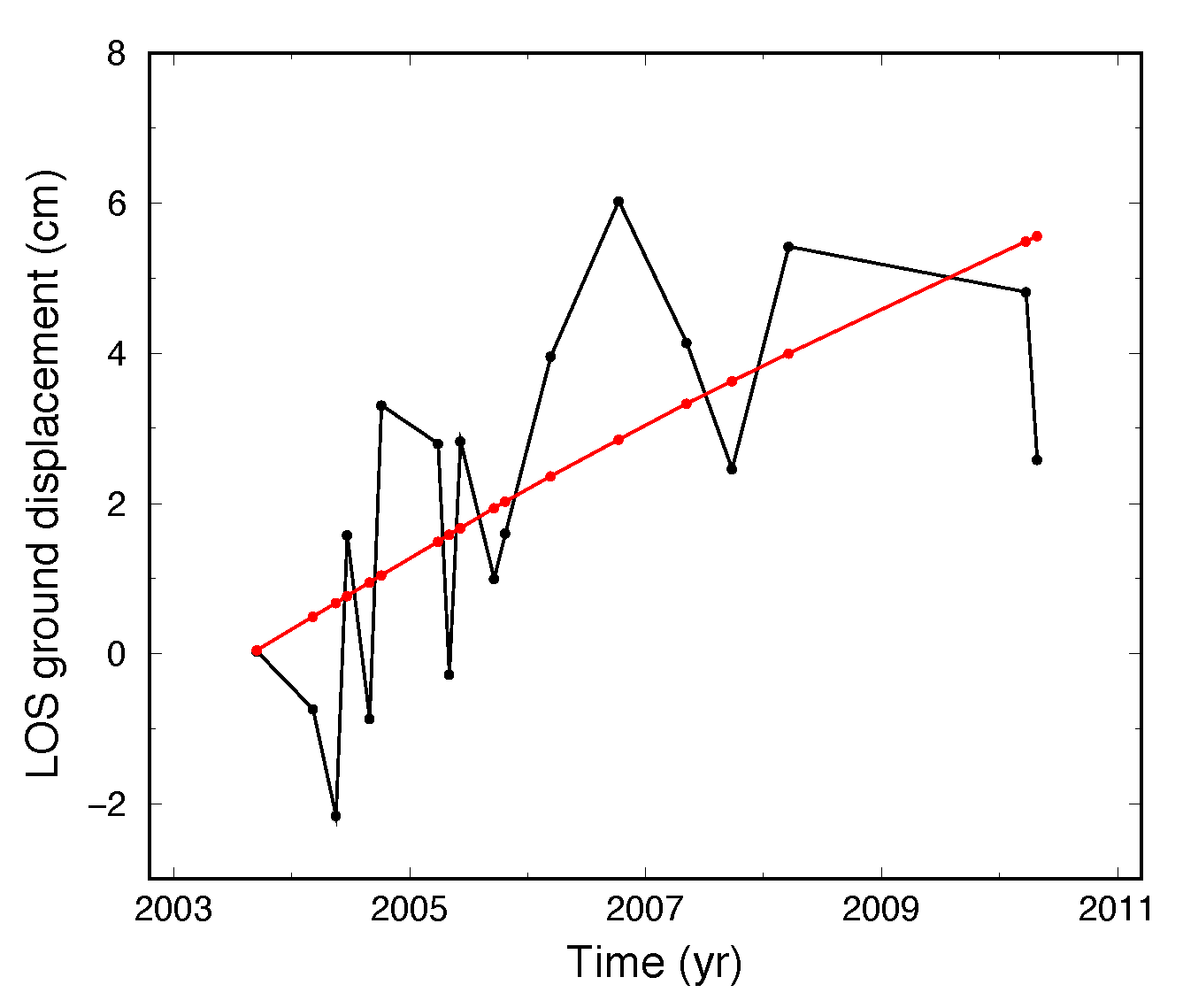}
\caption{\label{fig:ts} Example of ground displacement times series for a pixel located on the Anatolian plate (see the cross in Figure \ref{fig:geo_X} for location). Black and red lines represent the ground displacement between 2003 and 2011 without or with smoothing, respectively.\label{fig:disp}  }
\end{figure}

Figure \ref{fig:ts} shows the time series of the ground displacement in the satellite  line of sight (LOS) with or without smoothing in the inversion. We see that after smoothing, ground displacement rate becomes steady. However, it does not mean that all the high frequency APS have been removed. Actually, if it seems so in time for one pixel, the resulting smoothed displacement rate may change from one pixel to another due to different APS conditions. Spatially, APS are thus not fully removed, but its amplitude clearly decreased.

\begin{figure}[h]
\includegraphics[width=\linewidth]{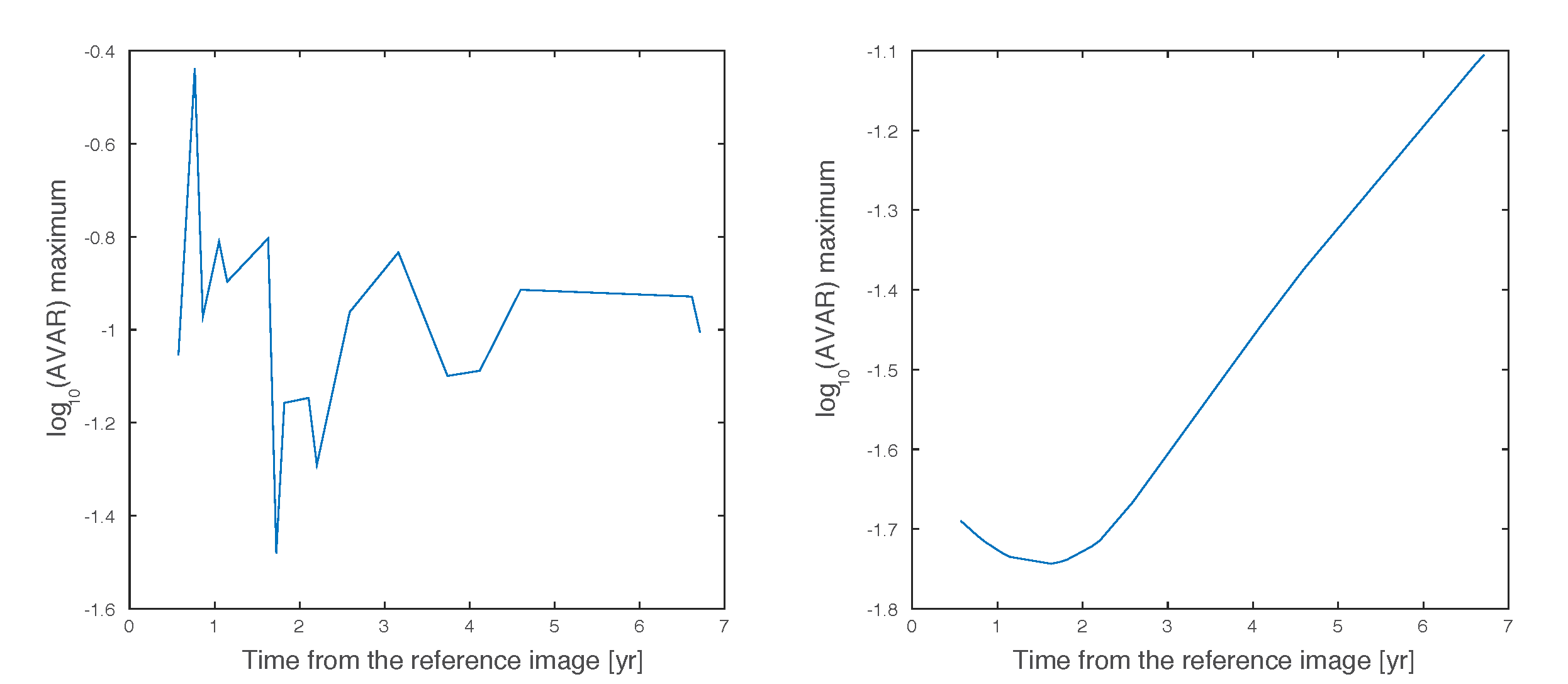}
\caption{\label{fig:avar_ts} Maximum of log$_{10}$(AVAR) value found at each incremental time step for ground displacement time series computed (a) without smoothing or (b) with smoothing. On (a), APS clearly dominates the log$_{10}$AVAR maximum value. On the contrary, on (b), after 2 years, when the tectonic signal starts to be strong enough, log$_{10}$AVAR maximum values increases proportionally with time, as for the model (Figure \ref{fig:avar_model1}c).}
\end{figure}

\begin{figure}[h]
\includegraphics[width=\linewidth]{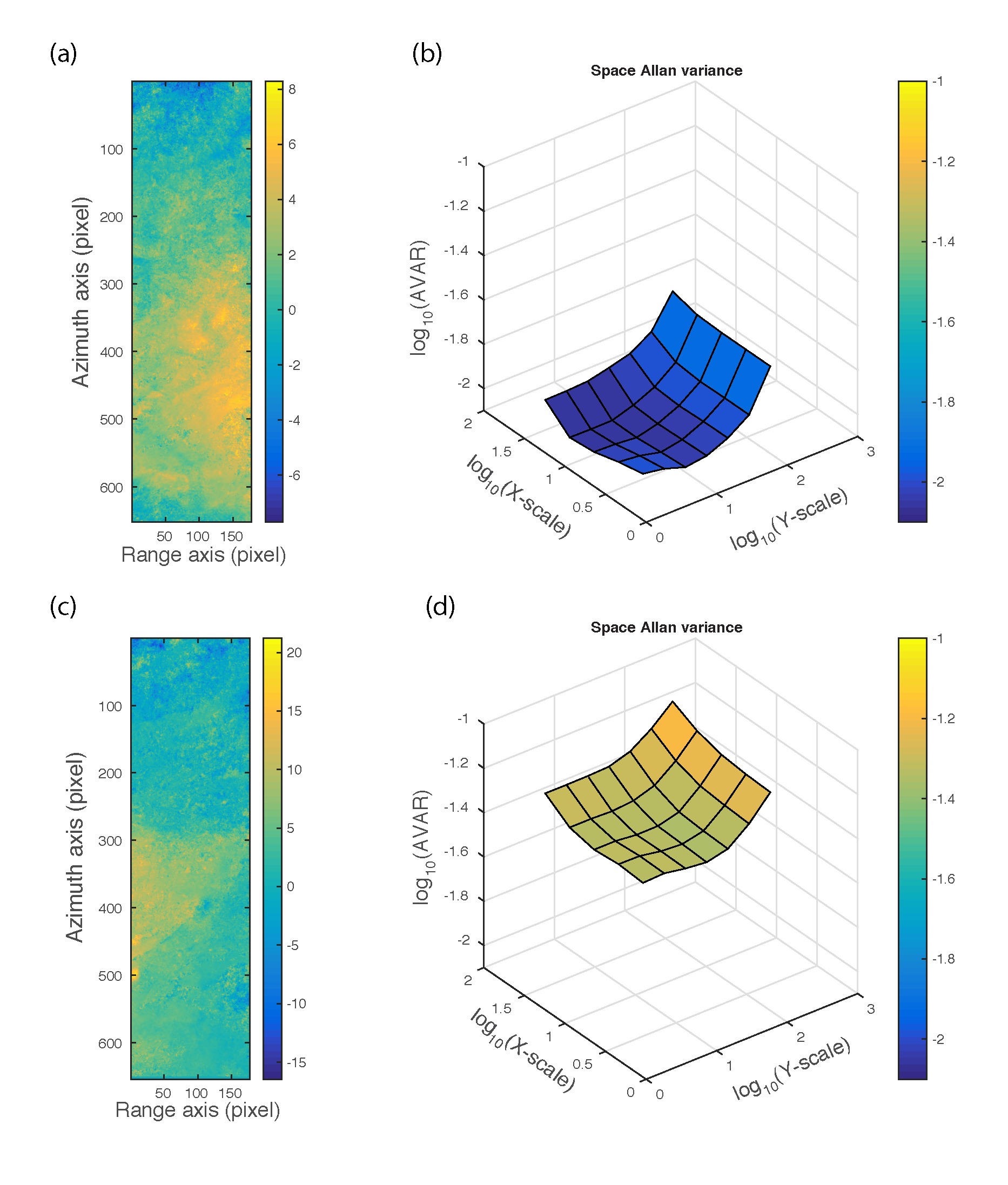}
\caption{\label{fig:avar_f} (a) Signal obtained after 1.05 years of observation  and (b) the corresponding space AVAR diagram. (c) Signal obtained over the whole time series and (d) the the corresponding space AVAR diagram.}
\end{figure}

As the tectonic signal contains long wavelength, to characterize it, we took the maximum value of the AVAR, $max_{av}$ obtained at largest scale factor. We did that for the non-smoothed and smoothed times series. Figure \ref{fig:avar_ts} shows the evolution $max_{av}$ with time (as a reminder, in times series, APS component amplitude is decorrelated from one to another SAR image, and depends on specific atmospheric conditions during the image acquisition, while tectonic signal grows steadily with time). We see that when no smoothing is applied,   no particular behavior of  $max_{av}$ shows up with time. It seems then that APS dominates the time series signal. On the other hand, for the smoothed time series,   $max_{av}$ increases linearly with time after 2 years. It seems then that the geological signal emerges from the noise after 2 years. Earlier, plate motion signal is in competition with the APS. Figure \ref{fig:avar_f} shows the time series  signal after 1 year, where the small geophysical signal is lost in the noise, and the final signal over the $\sim$7-year period of observation.

\subsection{Conclusion about InSAR application}

Space AVAR allows to distinguish between classical signals that one can deal with : white noise, random walk and others. In InSAR, different signals sum up and it can be difficult to observe the geophysical signal, typically here ground surface displacement due to plate tectonics, among  noise sources such as atmospheric delays or phase decorrelation. We saw that the phase decorrelation is easily identifiable and responds as white noise. Based on  theoretical, model atmospheric signature is also characteristic and differs significantly from other signals. However, in analyzing our data, we noted that atmospheric delays had additional large wavelength signal. This could be due to large topographic variations that would induce a correlated elevation component, not taking into account in the atmospheric model.  
Moreover, knowing the geophysical signal, we were able to compute its space AVAR signature which one feature is the larger AVAR values at large scale factors. As the tectonic signal is linear in time, it is easily verifiable if the time series has this characteristic. However, it is more difficult to assess the geophysical signal in space when it is mixed with different type of noise. In taking the space AVAR value for large scale factor,  we were then able to identify %
when the geophysical signal emerged distinctly from the noise.    

\section{Conclusion}
This study shows an original way to compute and use the Allan variance in the space domain.  It turns out that radial AVAR, called here Allan's hat, is the best way to obtain a stable AVAR solution, insensitive to the axis orientation. A fast algorithm has been elaborated in computing  the Fourier transform transfer function of the Allan's hat, that corresponds to a sum of two Bessel functions divided by the radial wave number (a sum of two Airy patterns).

We then applied this formalism to compute the space AVAR for InSAR data. It allowed to retrieve the main characteristics of noises that contaminate the ground displacement measurements obtained by computing the phase difference of two SAR images. We could clearly observe white noise due to phase decorrelation or quasi-planar AVAR features due to atmospheric delays. To distinguish between noises and geophysical signal, we extracted the AVAR value for large scale factors, where the geophysical signal is the strongest.  It showed great stability and revealed the ongoing  geophysical signal that emerged from noise after a period where noise and tectonic signal were of comparable amplitude. 

Through InSAR data, we showed that space AVAR can be useful to detect noise and characterize 2-D signals. It can thus be used to a multiple of other domains that deal with spatial data.  Compared to classical AVAR, space AVAR may give more complicated features to analyze, especially when occasionally the AVAR surface is bumpy and not planar, but seems very efficient for distinguishing noises from signal. Space AVAR could certainly be used more generally for image processing in other fields or even for other 2-D signals (surface analysis,\ldots). Moreover, the technique used here to go from 1-D to 2-D can also be extended to 3-D. We can imagine applications like volume temperature monitoring in a room, 3-D atmospheric evolution, etc. 

\section{Acknowledgements}
We wish to thank Michel Lenczner (Femto-ST, Besan\c{c}on, France) for his valuable help concerning the radial AVAR.

\bibliography{space_Avar.bib}
\end{document}